\documentstyle[psfig,11pt]{article}

\newcommand{\ps}{\not\!{p}}
\newcommand{\ks}{\not\!{k}}
\newcommand{\gam}{\gamma_{\mu}}

\newcommand{\gaf}{\gamma_5}
\newcommand{\eps}{\epsilon}

\newcommand{\ap}{\frac{\alpha}{4 \pi}}
\newcommand{\qi}{Q_i}
\newcommand{\qis}{Q_{i'}}
\newcommand{\qf}{Q_f}
\newcommand{\qfs}{Q_{f'}}
\newcommand{\pii}{p_i}
\newcommand{\pis}{p_{i'}}
\newcommand{\pf}{p_f}
\newcommand{\pfs}{p_{f'}}
\newcommand{\mi}{m_i}
\newcommand{\mis}{m_{i'}}
\newcommand{\mf}{m_f}
\newcommand{\mfs}{m_{f'}}
\newcommand{\ds}{\Delta_s}
\newcommand{\lo}{\log\left(\frac}
\newcommand{\lot}{\log^2\left(\frac}
\newcommand{\dw}{\Delta_W}
\newcommand{\dmw}{\Delta_{M_W}}
\newcommand{\dmz}{\Delta_{M_Z}}
\newcommand{\fyfs}{F_{Y\!F\!S}}
\newcommand{\fyfst}{\tilde{F}_{Y\!F\!S}}
\newcommand{\lw}{{\cal L}_W}

\begin{document}

\begin{titlepage}

\evensidemargin 0.7cm

\vspace{1cm}

\begin{flushright}
Fermilab-Pub-96/094-T \\
KA-TP-12-1996
\end{flushright}

\vspace{1cm}

\begin{center}
{\large \bf ELECTROWEAK RADIATIVE CORRECTIONS\\ \vspace{0.4cm}
TO RESONANT CHARGED GAUGE BOSON PRODUCTION} \\

\vspace{1.5cm}

{\rm DOREEN WACKEROTH}\footnote{E-Mail: dow@fnth09.fnal.gov}
 \\ \vspace{0.2cm}
{\em Fermi National Accelerator Laboratory} \\
{\em P.O. Box 500, Batavia, IL 60510, U.S.A.} \\

\vspace{1cm}

{\rm WOLFGANG HOLLIK} \\ \vspace{0.2cm}
{\em Institut f\"ur Theoretische Physik,
Universit\"at Karlsruhe} \\ {\em D-76128 Karlsruhe, Germany } \\

\vspace{4cm}

{\bf Abstract}	
\end{center}
\vspace{0.5cm}
The electroweak ${\cal O}(\alpha)$ contribution to 
the resonant single $W$ production in a general 4-fermion process
is discussed with particular emphasis on a gauge invariant 
decomposition into a QED-like and weak part.
The cross section in the vicinity of the resonance 
can be represented in terms of a convolution of a `hard'
Breit-Wigner-cross section, comprising
the $(m_t,M_H)$-dependent weak 1-loop corrections, with an
universal radiator function.
The numerical impact of the various contributions 
on the $W$ line shape are discussed, together with the concepts
of $s$-dependent and constant width approach. Analytic formulae for
the $W$ decay width are also provided 
including the 1-loop electroweak and QCD corrections.\\

\vspace{1cm}
\noindent
PACS: 12.15.Lk, 13.10.$+$q

\end{titlepage}

\section{Introduction}
\setcounter{equation}{0}\setcounter{footnote}{0}

Future experiments at LEP and the Tevatron will access 
sectors of the Minimal Standard Model (MSM) \cite{sm} yet unchallenged:
the Yang-Mills structure of gauge boson self couplings and mass
generation by the concept of spontaneous symmetry breaking \cite{hig}.
With LEP II operating above the threshold 
for $W$ pair production, for the first time 
a precise direct measurement of the triple gauge boson
coupling $(\gamma,Z)W^+W^-$ can be performed, allowing to test 
the non-Abelian structure of the MSM \cite{tgc}.
Moreover, our 
current knowledge of the $W$ boson mass (world average value \cite{mwav})
\[ M_W = 80.33 \pm 0.15 \; \mbox{GeV}\]
will be improved up to an uncertainty in the range of
30-50 MeV at LEP II \cite{mwac} and 20-30 MeV at the Tevatron upgrade
\cite{mwacc}.

Thus, in order to meet the precision of these future experiments
the knowledge of the observed cross sections 
beyond leading order perturbation theory is crucial.

The $W$ pair production cross section in the limit of stable $W$ bosons
beyond leading order is already known \cite{wwcross}, but not sufficient
at CM energies only a few $W$ boson decay widths above the
threshold. In the course of the calculation of the corrections to the realistic
scenario at LEP II with the subsequent decay of the $W$ bosons into fermions:
$e^+e^-\rightarrow W^+W^- \rightarrow 4 f$
the following problems arise:
\begin{itemize}
\begin{enumerate}
\item
the production and decay of $W$ bosons in the vicinity of
the threshold, where two energetically strongly varying phenomena occur:
the resonant cross section at $\sqrt{s_{\pm}}=M_W$ 
($s_{\pm}$: invariant masses of the outgoing fermion pairs)
and its increase at the threshold $\sqrt{s}=2 M_W$;
\item
the consistent treatment of unstable charged gauge bosons
within perturbation theory,
which involves infra-red singular interactions with real and 
virtual photons.
\end{enumerate}
\end{itemize} 
At present, there exists no complete calculation of the electroweak
${\cal O}(\alpha)$ contribution 
to the off-shell $W$ pair production cross section:
explicit results have been derived only for parts of the 
photonic corrections.
An overview on the present knowledge of the 
off-shell $W$ pair production beyond leading order and the
concessions to the consistency of the theory in order to gain it 
is given in \cite{wwcross}. 

The idea of this paper is to contribute to the description 
of charged unstable gauge bosons beyond leading order 
perturbation theory by studying 
the second problem separately and discussing 
the electroweak ${\cal O}(\alpha)$ contribution 
to the resonant single $W$ production in a 4-fermion process:
$i i' \rightarrow W^+ \rightarrow f f'$.
It appears as part of the $t$-channel
$W$ pair production process and its better understanding
can show a way to an improved description of
the off-shell $W$ pair production. Moreover, it represents the
$W$ production process via the Drell-Yan-mechanism at the Tevatron
and thus, in view of 
the future improved $W$ mass measurement at hadron colliders,
requires a careful treatment beyond lowest order in perturbation theory.

The discussion of the electroweak radiative corrections to the
$W$ production in the vicinity of the resonance
is guided by the successful treatment
of the $Z$ line shape beyond leading order \cite{cern},
which has been precisely measured at LEP I and SLC \cite{topfit}.
In contrary to the $Z$ resonance the electroweak radiative corrections to the
resonant $W$ production can not be naturally subdivided into a gauge invariant
photonic and non-photonic part. 
A separated treatment is motivated by the following reasons:
\begin{itemize}
\item
Usually, the photon contribution depends on cuts imposed
on the photon phase space and thus is dependent on the experimental setup.
\item
The enhancement of the fine structure constant $\alpha$ due
to large logarithms $\log(s/m^2)$ arising in connection with
infra-red (IR) and collinear singularities requires either 
the consideration of higher orders in perturbation theory or
the performance of a suitable resummation procedure.
\item
The interesting model-specific contributions are
contained in the non-photonic sector.
\end{itemize}
Therefore, in analogy to the description of the $Z$ resonance, we seek 
a consistent gauge invariant representation 
of the resonant $W$ production cross section
of the inclusive process $i i'\rightarrow W^+ \rightarrow f f' X$
with $X=$photons as a convolution integral of the following form \cite{bcern}:
\begin{equation}
\label{falts}
\sigma(s) = \frac{1}{s}\; \int_{s_0 = 4 m_f^2}^s 
d s' \; G(z) \; \sigma_w(s') \; .
\end{equation}
The shift of the invariant mass squared $s'= z s$ of the final state fermions
is due to initial state photon emission, which is described by
the universal radiator function $G(z)$. The latter also takes into account 
the possibility of multiple soft photon emission. The 
model dependent `hard' cross section
$\sigma_w(s)$ has a Breit-Wigner form. In next-to-leading order 
perturbation theory $\sigma_w(s)$ comprises the weak ($m_t, M_H$)-
dependent ${\cal O}(\alpha)$ contribution. 

The paper is organised as follows: 

In Sec.~3, after recalling the Born-cross section and the
tree level $W$ width (Sec.~2), we
concentrate on the gauge invariant separation of the electroweak 
${\cal O}(\alpha)$ contribution to the $W$ production 
into a QED-like and (modified) weak contribution. 
Our starting point is a thorough perturbative treatment of the
1-loop corrections to the lowest order matrix element.
For checking the cancellation of the 
unphysical gauge parameter dependence the calculation
is performed in $R_{\xi}$-gauge.
The application of the
procedure developed in \cite{yfs}
in order to extract a gauge invariant
multiplicative factor to the Born-cross section from the
IR-singular photon contribution leads to
QED-like form factors describing the initial, final state
and interference contribution, separately U(1) gauge invariant. 
In the resonance region, the remaining 
interference term can be absorbed into a
modified weak contribution, which then also factorises.
After performing an equivalent discussion of the
electroweak ${\cal O}(\alpha)$ contribution to the partial $W$ width 
($\rightarrow$ App.~B), the numerator of the Breit-Wigner
can be represented as a product 
of $W$ partial widths describing the
$W$ production and decay, respectively. 
At the end of Sec.~3, after a detailed discussion of the
QED-form factors and the modified weak contribution, we present
the cross section including the electroweak radiative corrections to 
the $W$ production in the vicinity of the resonance 
in terms of the convolution integral given by Eq.~\ref{falts}.
After a brief summary (Sec.~4) we provide
numerical results for the various
contributions in Eq.~\ref{falts} accompanied by a
numerical discussion of the $W$ decay width including 1-loop
electroweak corrections and QCD corrections (Sec.~5).

In App.~A, we discuss the aspect of gauge invariance in
the description of an unstable charged gauge boson 
beyond leading order from a more fundamental point of view.
The problem of a consistent
description of an unstable particle together with a 
definition of mass and width,
which meets the requirement of gauge invariance
order by order in perturbation theory, already had to be solved 
in the context of the precision measurements at the $Z$ resonance.
There, two approaches have been discussed: 
the S-Matrix theory inspired ansatz and the 
quantum field theoretical approach, yielding
a description with constant and $s$-dependent width, respectively.
The resulting prescriptions derived for the
$Z$ resonance need to be tested with regard to consistency and applicability
to the $W$ resonance, facing the additional difficulty of having 
IR-singular interactions of the $W$ boson with
virtual or real photons. At the end of App.~A the corresponding
prescriptions for the case of a charged vector boson resonance 
will be provided, especially, a transformation will be derived,
which connects both descriptions and enables the consideration
of an $s$-dependent $W$ width in Eq.~\ref{falts} in an easy way.
In the remaining appendices the explicit expressions for
the electroweak ${\cal O}(\alpha)$ contribution to the $W$ production and
$W$ width are provided and some details of the calculation are shown.

\section{$W$ production and $W$ width in leading order}
\setcounter{equation}{0}\setcounter{footnote}{0}

The decay width of a $W$ boson into quarks or leptons
in leading order perturbation theory, which
is graphical represented by the decay process in Fig.~1 (with $q^2=M_W^2$),
is given by \cite{width}
\begin{eqnarray}
\label{gwbmass}
\Gamma^{(0)}_{W\rightarrow f\!f'}
& = & \frac{\alpha M_W}{12 s_w^2} \; N_c^f \; |V_{f\!f'}|^2 \; 
\frac{1}{M_W^2}\; \sqrt{(M_W^2-(\mf+\mfs)^2)(M_W^2-(\mf-\mfs)^2)} \times
\nonumber\\
& & 
\left[1-\frac{\mf^2+\mfs^2}{2M_W^2}-\frac{(\mf^2-\mfs^2)^2}{2M_W^4}\right]\; ,
\end{eqnarray}
where $\alpha$ and $s_w$ denote the fine structure constant and
the sine of the Weinberg-angle, respectively. The quark mixing
is taken into account by the Kobayashi-Maskawa-matrix elements 
$V_{ij}$ \cite{koba} with $V_{i\!j}=\delta_{i\!j}$ for leptons. 
$N_c^f$ denotes the colour factor with $N_c^{f=l,q}=1,3$.
By using the leading order relation for the Fermi-constant 
$G_{\mu}$ (measured in the $\mu$-decay)
\begin{equation}
\label{drborn}
M_W^2=\frac{\pi\alpha}{\sqrt{2} G_{\mu} s_w^2}
\end{equation}
the partial $W$ width in the limit of massless decay products 
turns to
\begin{equation}
\label{gmudar}
\overline{\Gamma}^{(0)}_{W\rightarrow f\!f'}
= \frac{\sqrt{2} G_{\mu}  M_W^3}{12 \pi} \; N_c^f \; |V_{f\!f'}|^2 \; .
\end{equation}
This $G_{\mu}$-representation has the advantage to being independent of $s_w$.
The total width results from the summation of the partial decay widths 
into all fermionic final states compatible with energy momentum conservation
\begin{equation}
\label{gammatot}
\Gamma^{(0)}_W = \sum_{(f,f')} \Gamma^{(0)}_{W\rightarrow f\!f'}\; .
\end{equation}

\begin{figure}
\label{tree}
\hspace{4cm}
\psfig{figure=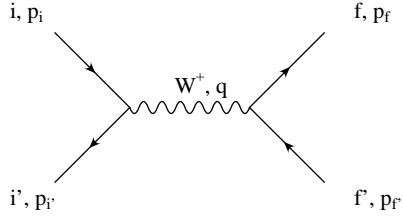}
\caption{$W$ production in the 4-fermion process at leading order}
\end{figure}

The production of a $W$ boson in a 4-fermion process in leading order
perturbation theory
is graphical represented by the Feynman-diagram
shown in Fig.~1.
We choose the Mandelstam variables 
\begin{eqnarray}
s &= &q^2 = (\pf+\pfs)^2 = (\pii+\pis)^2 
\nonumber\\
t &=& (\pf-\pii)^2 = (\pfs-\pis)^2 = -\frac{s}{2}(1-\cos\theta)
\nonumber\\
u &= &(\pf-\pis)^2 = (\pfs-\pii)^2 \; .
\end{eqnarray}
$\theta$ denotes the scattering angle of the outgoing fermion $f$
with respect to $\vec{p}_i$.
The differential cross section for this two-particle scattering
process can be written as follows:
\begin{equation}
\label{dwqpb}
\frac{d{\sigma}}{d t} =
\frac{1}{16\pi s^2} \overline{\sum} |{\cal M}|^2(s,t)  
\end{equation}
with the matrix element squared and averaged (summed) over the 
initial (and final) state spin and colour degrees of freedom. 
With the momentum assignment of Fig.~1 the  
Born-matrix element of the $W$ production in the limit of
massless external fermions yields as follows:
\begin{equation}
\label{matborn}
{\cal M}^{(0)} = i \frac{\pi \alpha}{2 s_w^2} V_{i i'} V_{f\!f'}
\;\frac{\overline u_f(\pf,s_f)
\gam (1-\gaf) v_{f'}(\pfs,s_{f'}) \overline v_{i'}(\pis,s_{i'})
\gamma^{\mu} (1-\gaf)u_i(\pii,s_i)}{s-M_W^2}\; . 
\end{equation}
In the vicinity of the resonance the Dyson-resummed propagator has to be used
(Eq.~\ref{prop}),
so that the differential Born-cross section of the resonant $W$ production has
Breit-Wigner-form
\begin{equation}
\label{dwqborn}
\frac{d \sigma^{(0)}(s,t)}{d t} = \frac{\pi \alpha^2}{s_w^4 \, s^2}
\; |V_{i i'}|^2 \, |V_{f\!f'}|^2 \, \frac{N_c^f}{N_c^i}  \;
[\frac{1}{2};\frac{1}{4}] \; \frac{(s+t)^2}
{[(s-M_W^2)^2+M_W^2 (\Gamma_W^{(0)})^2]} \; .
\end{equation}
The square bracket takes into account, that
for the case of incoming leptons the spin average 
yields only a factor $1/2$, since the neutrino is a 
purely left-handed particle, whereas the average over quark spins 
leads to a factor $1/4$. 
After performing the integration over the Mandelstam
variable $t$ ($-s<t<0$) the total cross section
of the resonant $W$ production in leading order
perturbation theory yields
\begin{equation}
\label{wborn}
\sigma^{(0)}(s) = \frac{\pi \alpha^2}{3\; s_w^4}
\; |V_{i i'}|^2 \, |V_{f\!f'}|^2 \, \frac{N_c^f}{N_c^i} \;
[\frac{1}{2};\frac{1}{4}] \; \frac{s}
{[(s-M_W^2)^2+M_W^2 (\Gamma_W^{(0)})^2]} \; ,
\end{equation}
which in $G_{\mu}$-representation is given by
\begin{equation}
\label{wbgmu}
\overline{\sigma}^{(0)}(s) = \frac{2 G_{\mu}^2 \; M_W^4}{3 \pi}
\; |V_{i i'}|^2 \, |V_{f\!f'}|^2 \, \frac{N_c^f}{N_c^i} \;
[\frac{1}{2};\frac{1}{4}] \; \frac{s}
{[(s-M_W^2)^2+M_W^2 (\overline{\Gamma}_W^{(0)})^2]} \; .
\end{equation}

\begin{figure}[htb]
\label{virt}
\mbox{\psfig{figure=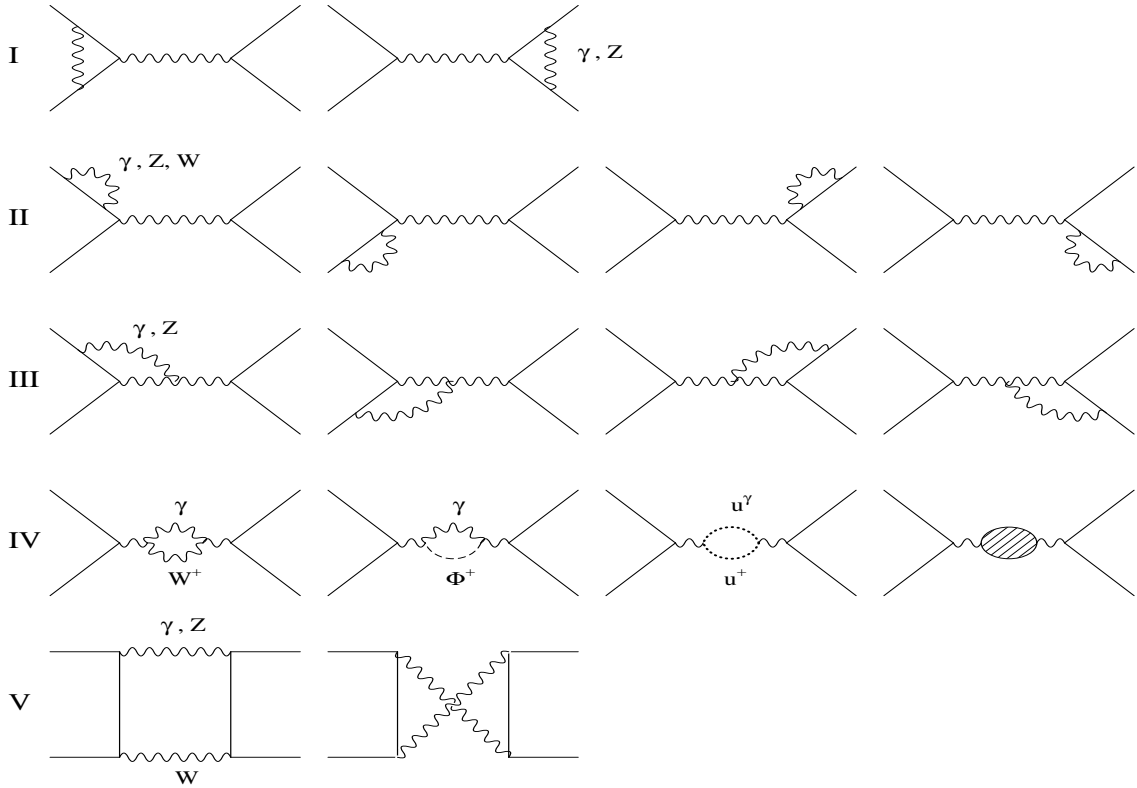,width=15cm,height=10.5cm}}
\caption{1-loop corrections to the $W$ production in the 4-fermion process
($\Phi^+$: Higgs-ghost, $u^+,u^{\gamma}$: Faddeev-Popov-ghosts;
the non-photonic contribution
to the $W$ self energy is symbolised by the shaded loop; an explicit
representation can be found in [22], e.g.)}
\end{figure}

\begin{figure}[htb]
\psfig{figure=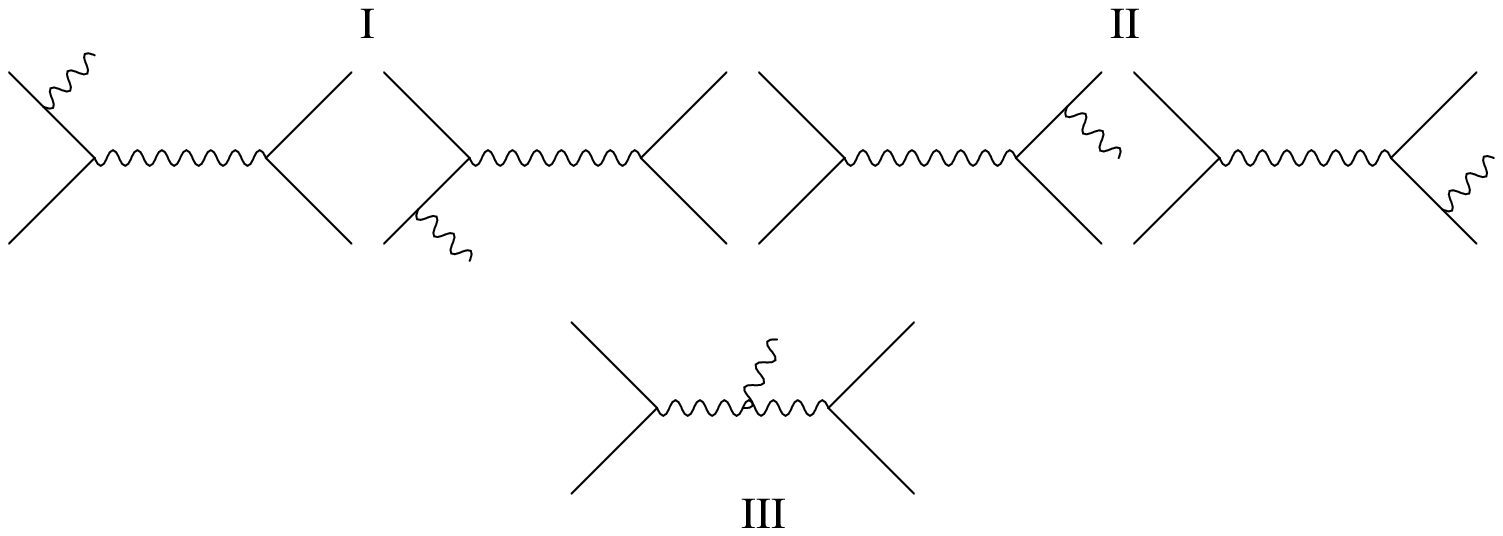}
\label{brp}
\caption{Real photon contribution in ${\cal O}(\alpha)$ to the
$W$ production in the 4-fermion process}
\end{figure}

\section{Electroweak radiative corrections in ${\cal O}(\alpha)$ to the 
$W$ production}
\setcounter{equation}{0}\setcounter{footnote}{0}

As motivated in the introduction,
our aim is to provide a consistent description of the $W$ resonance
beyond lowest order perturbation theory in form
of a convolution integral given by Eq.~\ref{falts}. 
To this end, a gauge invariant separation of the
electroweak radiative corrections under consideration into a QED-like and
weak contribution is required.

The starting point is a
perturbative treatment of the $W$ production in the
4-fermion process in ${\cal O}(\alpha^3)$.
The electroweak ${\cal O}(\alpha)$ contributions under consideration are
schematically represented by the Feynman-diagrams depicted 
in Fig.~2 and Fig.~3.
The virtual electroweak contribution, shown in Fig.~2, consists of vertex 
corrections due to photon and $Z$ boson exchange
(diagram I,III), self energy insertions
to the external fermions (diagram II), the $WZ$ and $W\gamma$ box diagrams
(diagram V) and
the $W$ self energy contribution (diagram IV).
Since the calculation is performed in $R_{\xi}$-gauge,
the latter also involves Higgs- and Faddeev-Popov-ghosts.
After renormalisation (here we work in the on-shell
scheme \cite{spiess}) the virtual contribution 
can be described by means of a gauge parameter ($\xi_i$,$i=\gamma,Z,W$)
independent, UV-finite, 
but IR-singular, form factor $\hat F_{virt.}(s,t)$ 
( $\hat{}$ denotes renormalised quantities) multiplying the
Born-cross section given by Eq.~\ref{dwqborn}.
When taking into account the real soft photon emission
(photon momentum $|\vec k|<\Delta E\ll \sqrt{s}$), shown in Fig.~3,
which can also be done in form of a multiplicative
IR-singular factor $F^s_{B\!R}(s,t)$,
the IR-singularities cancel as expected \cite{br}.
Finally, the $W$ production in ${\cal O}(\alpha^3)$
in a 4-fermion process can be described by
\begin{equation}
\label{notres}
\frac{d \sigma^{(0+1)}(s,t)}{d t} = 
\frac{d \sigma^{(0)}(s,t)}{d t} \; [1+2\;{\cal R}\!e \; \hat F_{virt.}(s,t) 
+F^s_{B\!R}(s,t)]\; ,
\end{equation}
where the explicit expressions for the contributions to
$\hat F_{virt.}(s,t)$ and $F^s_{B\!R}(s,t)$ of Eq.~\ref{virtsep}
and \ref{brinv}, resp., are provided in App.~D.
For the special choice $\xi_i=1$ the 
electroweak 1-loop corrections described by $\hat F_{virt.}(s,t)$
can also be found in \cite{spiess}.
The remaining photon phase space integration 
over the hard photon region is done in App.~E.

In the following,
we concentrate on the virtual electroweak contribution
and discuss the photon contribution $F_{\gamma}$
separately from the non-photonic pure weak contribution
$F_{weak}$
\begin{equation}
\label{virtsep}
\hat F_{virt.}(s,t) = (F_{\gamma}+F_{weak})(s,t)  \; .
\end{equation}
The virtual photon contribution comprises all Feynman-diagrams
in Fig.~2 involving a photon, where the photonic correction to the
$W$ self energy is 
explicitly represented by the first three diagrams of the subset IV.
In contrary to the $Z$ production, these Feynman-diagrams do not
build a gauge invariant subset and thus
$F_{\gamma}(s,t)$ and $F_{weak}(s,t)$
are UV-divergent and gauge parameter dependent.

Since, finally, we are only interested in the cross section in the vicinity 
of the $W$ resonance,
we have a closer look on the resonance structure of the 
different contributions to the virtual corrections depicted in Fig.~2.
It turns out, that the $WZ$-box diagrams
can be neglected as a non-resonant contribution of higher order, so that 
in the vicinity of the $W$ resonance 
the pure weak contribution in next-to-leading order 
evaluated at $s=M_W^2$
\begin{equation}
\label{wres}
F_{weak}(M_W^2) = (F_{weak}^i+F_{weak}^f)(M_W^2) \; .
\end{equation}
is determined
according to the prescription given in App.~A (Eq.~\ref{rescon}).
The resulting form factors $F_{weak}^{(i;f)}(M_W^2)$ describe the non-photonic
1-loop corrections to the $W$ production and decay, respectively,
and are explicitly given by Eq.~\ref{purewf}.

Far more involved is the calculation of the photonic form factor 
$F_{\gamma}(s,t)$:
the non-factorisable $W\gamma$-box diagram is a resonant
contribution and has to be considered at the required level of accuracy,
the arising IR-singularities have to cancel
and logarithms of the form $\log(s-M_W^2)$, which diverge for $s\rightarrow
M_W^2$ (on-shell singularities), needed to be regularised in a gauge
invariant way, when approaching the resonance region.
In order to obtain a separation of the 1-loop corrections 
into a QED-like and weak contribution, 
we first extract gauge invariant form factors,
so-called YFS-form factors $\fyfst^a(s)$, from the 
IR-singular Feynman-diagrams I,II and V (Fig.~2), so that the 
virtual photon contribution can be written as follows:
\begin{equation}
\label{photz}
F_{\gamma}(s,t)= \sum_{a=initial,final,\atop interf.} 
\fyfst^a(s)+F_{\gamma}^{finite}(s,t) \; .
\end{equation}
These YFS-form factors together with the real photon contribution 
build IR-finite gauge invariant form factors $F_{Q\!E\!D}^a(s,t)$, which are
independent from the internal structure of the $W$ production
and thus can be interpreted as a QED-like correction.
For that, the bremsstrahlung contribution, shown in Fig.~3,
needs also to be represented by a separately conserved 
initial and final state current, which  
cannot be easily obtained due to the $\gamma W^+W^-$-coupling in diagram III.
The sum of the remaining IR-finite contribution $F_{\gamma}^{finite}(s,t)$,
a part of the QED-form factor  
describing the interference of initial and final state
bremsstrahlung $F_{Q\!E\!D}^{interf.}$
and the pure weak part $F_{weak}^{i,f}$
represents a form factor $\tilde F_{weak}^{i,f}$,
which is independent of the 
external fermions and thus can be interpreted as a 
modified weak contribution. For the sake of clearness,   
the characteristics of the electroweak corrections in ${\cal O}(\alpha)$
are summarised and the different steps, which lead to
a description of the $W$ resonance given by Eq.~\ref{falts}, are schematically
presented in Tab.~1.
\renewcommand{\arraystretch}{1.5}
\begin{table}[htb]\centering
\label{schema}
\begin{tabular}{|c|c|c|c|c|}\hline \hline
$F^s_{B\!R}(s,t)$ & 
\multicolumn{4}{|c|}{$\hat{F}_{virt.}(s,t)$} \\ \hline \hline
$F^s_{B\!R}(s,t)$ & 
\multicolumn{3}{|c|}{$F_{\gamma}(s,t)$} &
$F_{weak}(M_W^2)$ \\ \hline
Fig.~3: $I,I\!I,I\!I\!I$
 & \multicolumn{2}{|c|}{Fig.~2: 
$\underbrace{I \; I\!I}_{U\!V,\xi_i,I\!R}
\; \underbrace{V}_{\xi_i,I\!R,os}$} & 
$\underbrace{I\!I\!I \; I\!V}_{U\!V,\xi_i,os}$ &
$\underbrace{I\ldots I\!V}_{U\!V,\xi_i}$ \\ \hline
$\underbrace{F_{B\!R}^{initial}(s)}_{I\!R,os}
+\underbrace{F_{B\!R}^{final}(s)}_{I\!R}$ &
$\underbrace{\fyfst^{initial}(s)
+\fyfst^{final}(s)}_{I\!R}$ &
$\underbrace{F_{I,I\!I,V}^{finite}(s,t)}_{U\!V,\xi_i,(subtr.)}$ & 
$\underbrace{F_{I\!I\!I,I\!V}^{\gamma}(s)}_{U\!V,\xi_i,(subtr.)}$ & 
$\underbrace{F_{weak}^{i,f}(M_W^2)}_{U\!V,\xi_i}$ \\  
$+\underbrace{F_{B\!R}^{interf.}(s,t)}_{I\!R,os}$ &
$+\underbrace{\fyfst^{interf.}(s,t)}_{I\!R,os}$ & 
&  & \\ \hline
\multicolumn{1}{|c}{$\underbrace{(F_{Q\!E\!D}^{initial}}_{os}+
F_{Q\!E\!D}^{final})(s)$} &  $+
\underbrace{F_{Q\!E\!D}^{interf.}(s,t)}_{F_{Q\!E\!D}^{interf.}|_{log.}
+\delta_{v+s}^{interf.}}$ &
\multicolumn{2}{|c|}{$\underbrace{F_{\gamma}^{finite}(s,t)}_{U\!V,\xi_i}
$}
& $\underbrace{F_{weak}^{i,f}(M_W^2)}_{U\!V,\xi_i}$  \\ \hline \hline
$\underbrace{(F_{Q\!E\!D}^{initial}}_{os}+
F_{Q\!E\!D}^{final})(s)$ &
\multicolumn{4}{|c|}{$(\tilde{F}_{weak}^i
+\tilde{F}^f_{weak})(M_W^2)$} \\ 
$+F_{Q\!E\!D}^{interf.}|_{log.}(s,t)$ & \multicolumn{4}{|c|}{} \\ \hline \hline
\end{tabular}
\caption{Scheme to the extraction of a QED-form factor to the $W$ production
($U\!V,\xi_i,I\!R,os$ denote the UV-divergence, $\xi_i$-dependence,
IR-singularity and on-shell singularity, resp.; $(subtr.)$ 
is referred to a prescription concerning the on-shell singularities,
which will be given in detail in Sec.~3.1}
\end{table}

In the following, this briefly outlined method to
find a gauge invariant separation into QED-like and weak part,
where even the ${\cal O}(\alpha)$ contribution to the $W$
production and decay process are separately represented by gauge invariant 
form factors, is going to be performed in detail.
 
\subsection{The definition of a QED-form factor to the $W$ production}

In the context of a general treatment of IR-singularities occurring in 
QED, Yennie, Frautschi and Suura (YFS) \cite{yfs} gave a prescription how to
separate these singularities as a multiplicative gauge invariant factor to the
Born-cross section. The basis of the perturbative treatment \`{a} la YFS
is the observation that the singularities arise only in connection with
soft photons emitted by external particles. The cross section of this 
soft photon radiation (virtual or real) can be described as the
Born-cross section and factors, which only depend on 
the four momenta of the external particles
and not on the internal structure of the process under consideration. 
This enables the treatment of soft photon radiation, especially the
demonstration of the cancellation of the IR-singularities, 
to all orders in perturbation theory.
In the following, the YFS-method 
will be applied to the photonic 1-loop contributions to the
$W$ production and later on also to the $W$ width.
By the example of the photon exchange between the final state fermions
the extraction of the YFS-form factor $\fyfs(s,t)$ 
from the diagrams I,II and V in Fig.~2 
will be illustrated.
The IR- and UV-singularities arising in the course of the calculation
are made mathematically well-defined
by introducing a fictitious photon mass $\lambda$
and by dimensional regularisation \cite{dim}, respectively.
The external fermions are considered 
in the massless approximation unless they occur
in singular logarithms of the form
$\log(s/m^2)$, where a finite fermion mass has been retained.
The explicit expressions for the IR-singular and -finite
parts of the diagrams under consideration can be found in App.~D.1.

The application of the Feynman-rules of the electroweak MSM 
leads to the following expression for the photonic final state correction 
described by diagram I: \\

\begin{minipage}{7cm}
{\psfig{figure=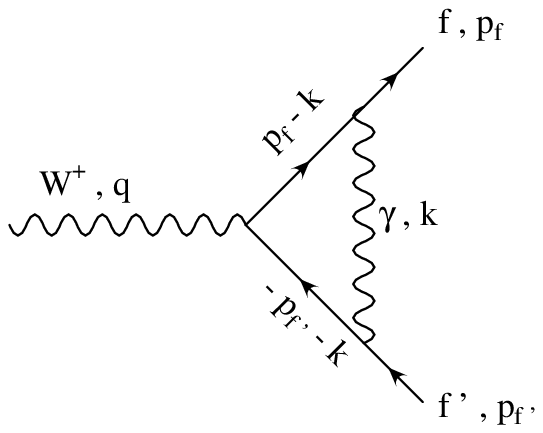}}
\end{minipage}\hfill
\parbox{8cm}{\begin{eqnarray}
:i\Lambda_{\mu}^{I,f}&=&i g_w \gam (1-\gaf)
\; [F_{I,f}^{I\!R}(s)+F_{I,f}^{finite}(s)]\nonumber\\
& & \left[g_w=\frac{e}{2 \sqrt{2} s_w}\right]
\end{eqnarray}}

\begin{eqnarray}
\label{diaone}
i\Lambda_{\mu}^{I,f}|_{\xi_i=1} & = & (-g_w) \; 4\pi\alpha \; \qf \qfs \int_D
\frac{\gamma_{\alpha} [\ps_f-\ks]\gam(1-\gaf)[\ps_{f'}+\ks]\gamma^{\alpha}}
{\underbrace{[k^2-\lambda^2]}_{D_{\lambda}}
\underbrace{[k^2-2(k\pf)]}_{D_f}\underbrace{[k^2+2 (k\pfs)]}_{D_{f'}}} \; .
\end{eqnarray}

Following the prescription given by YFS, the numerator
of the IR-singular Feynman-integral  
in Eq.~\ref{diaone} sandwiched in between the spinors
describing the final state fermions can be written as follows:
\begin{eqnarray}
\label{numer}
\mbox{numerator}& =& \overline{u}(\pf)
\gamma_{\alpha} [\ps_f-\ks]\gam(1-\gaf)[\ps_{f'}+\ks]\gamma^{\alpha}
v(\pfs) 
\nonumber\\
& = & \overline{u}(\pf)
[2 p_{f \alpha}-\gamma_{\alpha} \ks]\gam(1-\gaf)[2 \pfs^{\alpha}
+\ks\gamma^{\alpha}] v(\pfs)
\nonumber\\
& = &
\overline{u}(\pf)\gam(1-\gaf) v(\pfs)
(2 \pf-k)(2 \pfs+k)+ \mbox{terms}\propto \sigma^{\beta\alpha} k_{\beta} \; ,
\end{eqnarray}
where the following relations have been used:
\[\ks\gam=k_{\mu} I+\frac{1}{2}
[\ks,\gam]=k_{\mu} I-i\sigma_{\nu\mu} k^{\nu}\]
and
\[\overline{u}(\pf)\ps_f=m_f\overline{u}(\pf)=0 \; , \;
\ps_{f'} v(\pfs)=-\mfs v(\pfs)=0 \; .\]
The first term in Eq.~\ref{numer} leads to the IR-singular contribution
of diagram I, which will be part of the YFS-form factor
\begin{equation}
\label{ireins}
F_{I,f}^{I\!R}(s)  =  (i 4\pi\alpha) \qf \qfs 
\int_D \frac{(2 \pf-k)(2 \pfs+k)}{D_\lambda D_f D_{f'}} \; ,
\end{equation}
whereas the IR-finite 'magnetic' part contributes to 
$F_{\gamma}^{finite}(s,t)$ in Eq.~\ref{photz}.

The application of this procedure to the photonic self energy
insertions to the external final state fermions
and to the photonic box diagrams
leads to the following IR-singular form factors:
\begin{minipage}{6cm}
{\psfig{figure=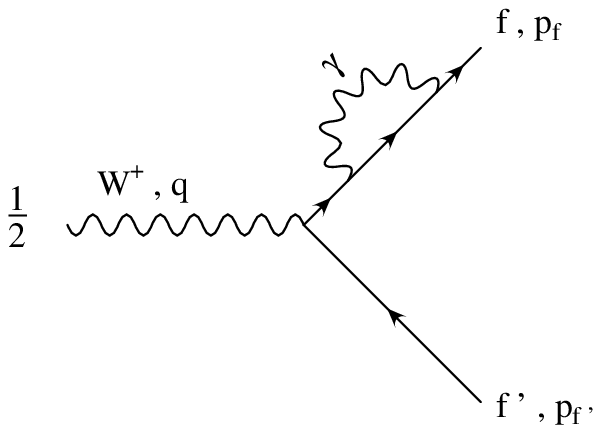}}
\end{minipage}
$+$ \hfill 
\begin{minipage}{6cm}
{\psfig{figure=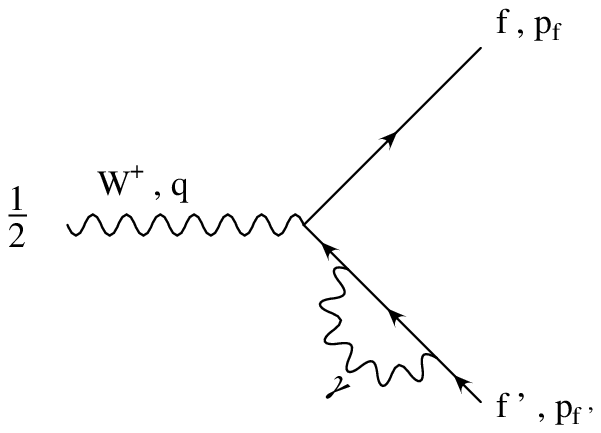}}
\end{minipage}
\begin{equation}
:\;  i\Lambda_{\mu}^{I\!I,f}  =  i g_w \gam (1-\gaf)
\; [F_{I\!I,f}^{I\!R}(s)+F_{I\!I,f}^{finite}(s)]
\end{equation}
with
\begin{equation}
\label{irzwei}
F_{I\!I,f}^{I\!R}(s) = \frac{1}{2} (i 4 \pi \alpha)
\left\{\qf^2 \int_D\frac{(2 \pf-k)^2}{D_\lambda D_f^2}
+\qfs^2 \int_D\frac{(2 \pfs+k)^2}{D_\lambda D_{f'}^2} \right\}
\end{equation}
and

\begin{minipage}{6cm}
{\psfig{figure=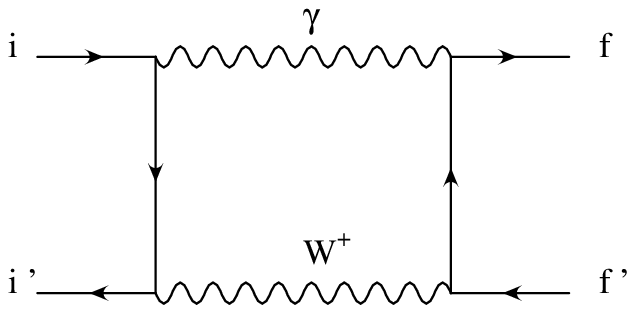}}
\end{minipage}
$+$ \hfill
\begin{minipage}{6cm}
{\psfig{figure=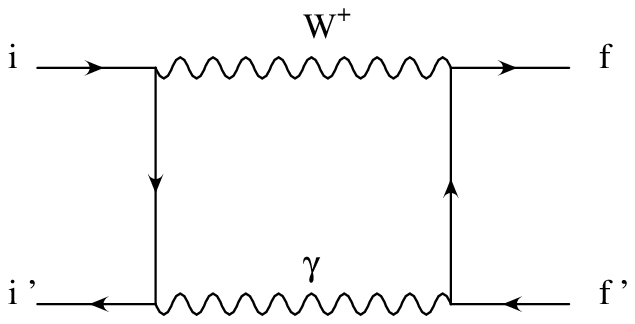}}
\end{minipage}
\begin{equation}
: \; i B^t(s,t)= i {\cal M}^{(0)}(s)
\; [F_{V,t}^{I\!R}+F_{V,t}^{finite}](s,t)
\end{equation}
with
\begin{equation}
\label{boxir}
F_{V,t}^{I\!R}(s,t)  =  -(i 4 \pi\alpha) \left\{\qi \qf
\int_D \frac{(2 \pf-k)(2 \pii-k)}{D_\lambda D_f D_i}
+\qis \qfs \int_D \frac{(2 \pfs+k)(2 \pis+k)}{D_\lambda D_{f'} D_{i'}}
\right\}  \; .
\end{equation}
The form factors describing the initial state vertex corrections
$F_{(I,I\!I),i}^{I\!R,finite}(s)$ can be derived from 
the final state ones by the substitution
\begin{equation}
\label{substi}
(\qf,\qfs,\mf,\mfs) \rightarrow (\qi,\qis,\mi,\mis) \; ,
\end{equation}
which in the following will be abbreviated by $(f,f') \rightarrow (i,i')$.
The $u$-channel form factors $F_{V,u}^{I\!R,finite}(s,u)$
describing the crossed box diagrams in Fig.~2
follow from $F_{V,t}^{I\!R,finite}(s,t)$ by the substitutions
\begin{equation}
\label{cross}
(i,f),(i',f')\rightarrow (i,f'),(i',f)  \; \mbox{and} \; t\rightarrow u 
\end{equation}
and, additionally, by multiplying with a global minus sign.
The Born-matrix element ${\cal M}^{(0)}$ is given by Eq.~\ref{matborn}.

These IR-singular form factors are extracted from the 
virtual photon contribution in such a way, that their 
sum has a structure similar to that of the amplitude describing
real (soft) photon radiation
\begin{eqnarray}
\label{yfsp}
\lefteqn{\fyfs(s,t) = 
\frac{1}{2} (i 4 \pi \alpha) \int_D \frac{1}{D_\lambda} \times}
\nonumber\\
& & \left[\overbrace{\frac{\qi (2 \pii-k)_{\rho}}{k^2-2k\pii} +
\frac{\qis (2 \pis+k)_{\rho}}{k^2+2k\pis}
-\frac{\qf (2 \pf-k)_{\rho}}{k^2-2k\pf}-
\frac{\qfs (2 \pfs+k)_{\rho}}{k^2+2k\pfs}}^{k^{\rho}{\cal J}_{\rho}=0}
\right]^2 \; . 
\end{eqnarray}
Thus, the $U(1)$-gauge invariance of the YFS-form factor
is guaranteed by the existence of a conserved current.  
The initial and final state contribution to the 
YFS-form factor, however,
distinguished by the corresponding charge quantum numbers
($\qi,\qis$ and $\qf,\qfs$) are {\em not} separately gauge invariant. 
Therefore, a `zero' will be added, so that the
YFS-form factor can be written as a sum of two separately 
conserved $U(1)$-currents, which describe the virtual photonic
correction to the $W$ production and decay process, respectively.  
\begin{eqnarray} 
\label{yfsfak}
\lefteqn{\fyfs(s,t)  =  \frac{1}{2} (i 4 \pi \alpha) \int_D
\frac{1}{D_\lambda} \times}
\nonumber\\
& & \left[\overbrace{\frac{\qi (2 \pii-k)_{\rho}}{D_i}+
\frac{\qis (2 \pis+k)_{\rho}}
{D_{i'}}-\frac{1}{2} \frac{(\qi-\qis) (2 q-k)_{\rho}}{(k^2-2 k q)}+
\frac{1}{2}\frac{(\qi-\qis) (2 q+k)_{\rho}}{(k^2+2 k q)}}^
{k^{\rho}{\cal J}^{initial}_{\rho} = 0}\right.
\nonumber\\
&+& \left. \overbrace{\frac{1}{2} \frac{(\qf-\qfs) 
(2 q-k)_{\rho}}{(k^2-2kq)}-\frac{1}{2} \frac{(\qf-\qfs) 
(2 q+k)_{\rho}}{(k^2+2kq)}
-\frac{\qf (2 \pf-k)_{\rho}}{D_f}-\frac{\qfs
(2 \pfs+k)_{\rho}}{D_{f'}}}^{k^{\rho}{\cal J}^{final}_{\rho} = 0}\right]^2
\nonumber\\
& =: & \fyfs^{initial}(s)+\fyfs^{final}(s)+\fyfs^{interf.}(s,t) \; .
\end{eqnarray}
This, at the first sight, arbitrary extension will receive its 
justification from the structure of the real photon contribution
and its interpretation in the course of the
corresponding discussion of the photon contribution to the $W$ width
(App.~B).
The explicit expressions for the gauge invariant form factors
after the evaluation of the loop integral in Eq.~\ref{yfsfak}
can be found in App.~D.1.
Before we deal with the real photon contribution
a closer inspection of the occurring mass singularities 
$\log(s/m^2)$ and logarithms of the form $\log(s-M_W^2)$
is needed. Since the occurrence of those singularities is a pure QED 
phenomenon, they build a gauge invariant subset  
\begin{equation}
[\fyfs^{initial,final,interf.}+F_{I,(f,i)}^{finite}+F_{V,(t,u)}^{finite}
+F_{I\!I\!I}]_{mass-sing.}
 =  \ap \sum_{k=i,i',f,f'} Q_k^2 \frac{1}{2}\lo{s}{m_k^2}\right)
\end{equation}
and ($\beta_{int.}$ of Eq.~\ref{betint})
\begin{equation}
[F_{V,(t,u)}^{finite}
+F_{I\!I\!I}+F_{I\!V}]_{on-shell-sing.}
 = \frac{1}{2} \; \beta_{int.}(s,t)\; \lo{M_W^2}{|s-M_W^2|}\right) \; ,
\end{equation}
which can be assigned to the initial state, final state and interference
YFS-form factors according to their structure and under the maintenance 
of gauge invariance.   
It has to be mentioned, that the sum of the IR-finite photon contributions
which are not included in the YFS-form factors develops a 
further QED-specific term  
\[ \ap \;  \sum_{k=i,i',f,f'} Q_k^2 \; ,\]
which thus can be absorbed in a modified YFS-form factor, as well.
Finally, the resulting modified YFS-form factors in
Eq.~\ref{photz} are connected to the original ones (Eq.~\ref{yfsfak})
as follows:
\begin{eqnarray}
\label{yfsmod}
\fyfst^{(initial;final)}& = & \fyfs^{(initial;final)}
-[\fyfs^{(initial,final)}]_{mass-sing.}
\nonumber\\
&+& \ap \sum_{k=(i,i');(f,f')} Q_k^2 \; 
\left[\frac{1}{2}\lo{s}{m_k^2}\right)-1\right]
\nonumber\\
\fyfst^{interf.} & = & \fyfs^{interf.} 
-[\fyfs^{interf.}]_{mass-sing.} 
+\frac{1}{2} \; \beta_{int.}(s,t)\; \lo{M_W^2}{|s-M_W^2|}\right) \; .
\end{eqnarray}
It is this modification which guarantees,
that the inclusive cross section including the hard final state photons
fulfills the KLN-theorem \cite{kln} and that the occurrence of the 
on-shell singularities is restricted to the initial state contribution. 

The last step to extract a QED-like form factor from the electroweak 
radiative corrections
to the $W$ production is to find a gauge invariant separation of the real
photon radiation into initial and final state contribution.
It turns out, that diagram III in Fig.~3 can be divided into one part, which 
develops the propagator structure of a initial state contribution and another
one, which can be assigned to the final state \cite{brb}  
\begin{eqnarray}
\mbox{diagram III} & : & \frac{1}{[q^2-M_W^2][(q-k)^2-M_W^2]} 
\nonumber\\
& = & \underbrace{\frac{1}{[(q-k)^2-M_W^2][2 kq]}}_{\leftarrow \mbox{initial
 state}}
-\underbrace{\frac{1}{[q^2-M_W^2][2 kq]}}_{\rightarrow \mbox{final state}}\; .
\end{eqnarray}
Using this separation the contribution of the real soft photons shown in 
Fig.~3 can be described by a multiplicative
factor being composed of separately conserved initial and final state
$U(1)$-currents   
\begin{eqnarray}
\label{brinv}
F^s_{B\!R}(s,t) & = & (-4 \pi \alpha) \int_{|\vec{k}|< 
\Delta E} \frac{d^3 k}{2 (2 \pi)^3 k^0}
\overbrace{\left|\frac{s-M_W^2}{(s-M_W^2-2 kq)} \left[
\frac{\qi \pii^{\rho}}{k\pii}-
\frac{\qis \pis^{\rho}}{k\pis}-\frac{(\qi-\qis) q^{\rho}}{kq}\right]\right.}^
{k_{\rho}{\cal J}_{initial}^{\rho} = 0}
\nonumber\\
&+&  \overbrace{\left. \frac{(\qf-\qfs) 
q^{\rho}}{kq}-\frac{\qf \pf^{\rho}}{k\pf}+\frac{\qfs
\pfs^{\rho}}{k\pfs}\right|^2}^{k_{\rho}{\cal J}_{final}^{\rho} = 0}
\nonumber\\
& =: &  F_{B\!R}^{initial}(s)+F_{B\!R}^{final}(s)+F_{B\!R}^{interf.}(s,t) \; .
\end{eqnarray}
There the impact of a photon radiated by a initial state fermion
to the $W$ propagator has been 
taken into account, as well. The explicit expression for the
gauge invariant form factors $F_{B\!R}^{a}(s,t)$
after performing the photon phase space
integration in the soft photon limit can be found in App.~D.3.

Finally, the QED-like form factors of the $W$ production, which correspond
to the QED-form factors describing the next-to-leading order 
photonic corrections to the $Z$ production, are determined by the 
YFS- and bremsstrahlung form factors derived above 
as follows:
\begin{equation}
\label{sumqed}
F_{Q\!E\!D}^a = 2\; {\cal R}\!e \;\fyfst^a
+F_{B\!R}^a \; \; \mbox{ with }\;  \; 
{\small a=initial,final,interf. } \; .
\end{equation}
Up to now, we only considered the radiation of soft photons,
since they develop IR-singularities, which have to be cancelled in the sum 
of the real and virtual contribution. 
In the following, it will be shown that 
in Eq.~\ref{sumqed} this cancellation works. Moreover, 
the radiation of hard photons will be considered
by performing the integration over the remaining
photon phase space: $k^0_{min}=\Delta E$ up to
$k^0_{max}=M_W/2$ as it is described in App.~E.
Since we are interested in the cross section of the 
$W$ production in the vicinity of the resonance, those
terms, which would vanish for $s \rightarrow M_W^2$, have been neglected.
Furthermore, the $W$ width will be introduced in order to cope with the arising
on-shell singular logarithms by the replacement  
\[s-M_W^2\stackrel{R}{\longrightarrow} \dw=s-M_W^2+i M_W \Gamma_W^{(0+1)} \;,\]
which can be done without spoiling the $U(1)$-current conservation as 
it can easily verified by Eq.~\ref{brinv}. The replacement of $\sigma^{(0)}(s)$
with $\tilde\sigma^{(0)}(s)$ (Eq.~\ref{wborn} with $\Gamma_W^{(0)}\rightarrow
\Gamma_W^{(0+1)}$) in the vicinity of the resonance follows the
prescription developed in App.~A.\\

\underline{{\bf The initial state QED-form factor}}:\\

\noindent
The gauge invariant QED-like contribution to the total cross section
in ${\cal O}(\alpha^3)$ in the vicinity of the $W$ resonance,
which has been extracted from the
virtual and real (soft) photonic initial state correction
to the $W$ production in the 4-fermion process, yields 
(Eqs.~\ref{fbrin},\ref{yfsfin} with Eq.~\ref{substi} and $\qi-\qis=1$)
\begin{eqnarray}
\label{sigsin}
\sigma_{i,v+s}^{(0+1)}(s)& = & \tilde{\sigma}^{(0)}(s) \;(1+
F_{Q\!E\!D}^{initial}(s))
\nonumber\\
& = &\tilde{\sigma}^{(0)}(s)\; \left\{1+
\beta_i(s) \left[
\lo{2 \Delta E }{\sqrt{s}}\left|\frac{\dw}{\dw-2\sqrt{s}\Delta E}
\right|\right)+\delta_p(s)\right]+2 \delta_{v+s}^i(s)\right\} 
\nonumber\\
& & \mbox{}
\end{eqnarray}
with
\begin{equation}
\label{beti}
\beta_i(s) = \frac{\alpha}{\pi} \left[\qi^2 \left(\lo{s}{\mi^2}\right)-1\right)
+\qis^2 \left(\lo{s}{\mis^2}\right)-1\right)-1\right] \; ,
\end{equation}
\begin{eqnarray}
\label{dvs}
\delta_{v+s}^i(s)& = &\ap \left\{\qi^2 \left[\frac{3}{2} \lo{s}{\mi^2}\right)
+\frac{\pi^2}{3}-2\right]+\qis^2 [i \rightarrow i']
+3+\frac{\pi^2}{12}\right\}
\end{eqnarray}
and the phase-shift of the resonance
\begin{equation}
\label{phaseshift}
\delta_p(s) = \frac{(s-M_W^2)}{M_W \Gamma_W^{(0+1)}} \left[
\arctan\left(\frac{s-M_W^2}{M_W \Gamma_W^{(0+1)}}\right)+
\arctan\left(\frac{2 \sqrt{s} \Delta E-s+M_W^2}{M_W \Gamma_W^{(0+1)}}\right)
\right] \; .
\end{equation}
This represents the main contribution to the entire 
electroweak 1-loop corrections due to the occurrence of 
large logarithms, for instance $\lo{\textstyle s}{\textstyle m_e^2}\right)
\approx 24$ for $s=M_W^2$. In the case of the $Z$ resonance a procedure
has been developed how to cope with those large contributions \cite{bcern}.
The achieved description of the initial state photon
contribution by the QED-form factor given by Eq.~\ref{sigsin}
enables now its application also to the $W$ resonance. 
For that purpose, the phase space integration over the hard photons
will be rewritten in accordance with Eq.~\ref{spekin}
by using $z=1-k=1-\frac{\textstyle{2k^0}}{\textstyle q^0}$ as follows:
\begin{eqnarray}
\label{sighin}
\sigma_{i,hard}^{(1)}(s)& = &\tilde{\sigma}^{(0)}(s)\; 
\int_{\eps}^1 \; dk \; \frac{|\dw|^2 (1-k)}{|\dw-sk|^2}
\; \left\{\beta_i(s) \frac{1}{k}+\frac{1}{2}\beta_i(s)\; (k-2)
+\frac{\alpha}{\pi}\frac{k}{6} \right\}
\nonumber\\
& = & \int_0^1 dz \; \theta(1-z-\eps) \; 
\tilde{\sigma}^{(0)}(sz) \left\{\frac{\beta_i(s)}{1-z}
+\tilde{\delta}_h(s) \right\} 
\end{eqnarray}
with $\eps=\frac{\textstyle{2\Delta E}}{\textstyle \sqrt{s}}$ and
$\tilde{\delta}_h$ is given by
\begin{equation}
\label{har}
\tilde{\delta}_h(s) = \frac{\alpha}{\pi}\frac{1-z}{6}
-\frac{1}{2}\; (1+z)\; \beta_i(s) \; .
\end{equation}
As it can easily be verified, the term $\propto 1/(1-z)$ of 
Eq.~\ref{sighin} cancels the $\Delta E$-dependence of
the soft QED-form factor. Thus, the cut-off parameter $\Delta E$ can be chosen 
to be so small that it can be neglected in Eq.~\ref{sigsin} as compared to
the $W$ width.  
As a consequence, the initial state bremsstrahlung
to the $W$ resonance can also be written
in form of a convolution integral
\begin{eqnarray}
\label{faltione}
\sigma^{(0+1)}_{i,s+h}(s) & = &\sigma^{(0+1)}_{i,v+s} +\sigma_{i,hard}^{(1)}
\nonumber\\
&= &\int_0^1 dz \; G^{(0+1)}(z)\; \tilde{\sigma}^{(0)}(sz) 
\end{eqnarray}
with the radiator function at 1-loop level
\begin{eqnarray}
G^{(0+1)}(z)& = & \delta(1-z)
\nonumber\\
&+& \delta(1-z) \left[\beta_i(s)\; \log(\eps)
+2\delta^i_{v+s}(s)\right]
\nonumber\\
& + &\theta(1-z-\eps) \; 
\left[\frac{\beta_i(s)}{1-z}+\tilde{\delta}_h(s) \right] \; .
\end{eqnarray}
This representation enables the consideration of 
the remaining electroweak 1-loop corrections 
and the effect of an $s$-dependent width in a simple way.
After performing the summation of the logarithms connected to the
soft photons to all orders in perturbation theory 
({\em soft photon exponentiation}) the convolution integral in
Eq.~\ref{faltione} reads as follows: 
\begin{equation}
\label{falti}
\sigma_{i,exp.}(s) = \int_0^1 dz \; G(z)\; 
\tilde{\sigma}^{(0)}(sz) \; ,
\end{equation}
with the radiator function in the exponentiated version
\begin{equation}
\label{radiator}
G(z) = \beta_i \; (1-z)^{\beta_i-1}\; (1+2 \delta_{v+s}^i)+\tilde{\delta}_h 
\; . 
\end{equation}
The calculation of the 
initial state bremsstrahlung at 2-loop level in the case of the
$Z$ resonance \cite{bcern}, either performed explicitly or 
by using the structure function method, has shown, that
the soft photon exponentiation together with
the remaining 1-loop contributions of the virtual and hard photons
represents the main part of the initial state bremsstrahlung.
A renormalisation group analysis \cite{been} confirms the method
of the summation of the leading logarithms arising in connection with 
the emission of soft photons ($\rightarrow$
Eq.~\ref{radiator}).\\

\underline{{\bf The {\em final state} QED-form factor}}:\\

\noindent
The gauge invariant QED-form factor describing the 
soft photons radiated by the final state fermions
is given by (Eqs.~\ref{yfsfin},\ref{fbrfin} and $\qf-\qfs=1$)
\begin{equation}
\label{qedfin}
F_{Q\!E\!D}^{final}(s) =  \beta_f(s) 
\lo{2 \Delta E}{\sqrt{s}}\right)+2 \; \delta_{v+s}^f(s)\; ,
\end{equation}
where $\beta_f(s)$ and $\delta_{v+s}^f(s)$ again can be derived from
the corresponding initial state expressions (Eqs.~\ref{beti},\ref{dvs})
by applying the substitutions $(i,i')\rightarrow (f,f')$.
After taking into account the radiation of hard photons
the so-defined soft photon contribution to the resonant $W$ production
cross section fulfills the KLN-theorem \cite{kln}
provided that no constraints on the invariant 
mass of the final state fermion pair will be imposed:
the mass singularities cancel out and finally 
a QED-form factor $\delta_{Q\!E\!D}^f$ remains multiplying the inclusive
total Born cross section 
\begin{equation}
\label{sigfin}
\sigma_{f,s+h}^{(0+1)}(s) = \tilde{\sigma}^{(0)}(s)\; 
(1+\delta_{Q\!E\!D}^f) \; ,
\end{equation}
which has the following form:
\begin{equation}
\label{qedfind}
\delta_{Q\!E\!D}^f = 
\frac{\alpha}{\pi}\; 
\left[\frac{3}{8}\; (\qf^2+\qfs^2)+\frac{7}{3}
+\frac{\pi^2}{24}\right]  
\stackrel{f,f'=\nu,l}\approx \;  0.0072 \; .
\end{equation}
Thus, as in the $Z$ resonance case,
this small effect of the final state bremsstrahlung 
can be taken into account by attaching a multiplicative factor
to the convolution integral in Eq.~\ref{falti}
\[\tilde{\sigma}^{(0)}(s)\;  \rightarrow \;
\tilde{\sigma}^{(0)}(s)\;   (1+\delta_{Q\!E\!D}^f)  \; .\]
\underline{{\bf The interference contribution}}:\\

\noindent
The interference of initial and final state soft bremsstrahlung  
leads to the following QED-form factor (Eqs.~\ref{yfsint},\ref{fbrint} with
$\qi-\qis=\qf-\qfs=1$):
\begin{equation}
\label{finterf}
F_{Q\!E\!D}^{interf.}(s,t)  = \underbrace{ \beta_{int.}(s,t)
\lo{2 \Delta E}{\sqrt{s}}\frac{M_W^2}{|\dw-2 \sqrt{s} \Delta E|}
\right)}_{\rightarrow 0 \; \mbox{for}\;  s=M_W^2,\Delta E \gg 
\frac{\Gamma_W}{2}}+2 \; \delta_{v+s}^{interf.}(s,t)
\end{equation}
with
\begin{equation}
\label{betint}
\beta_{int.}(s,t) = \frac{\alpha}{\pi}\left[(\qi\qf+\qis \qfs)\lo{t^2}{s^2}
\right)-(\qi\qfs+\qis \qf)\lo{u^2}{s^2}\right)+2\right]
\end{equation}
and
\begin{eqnarray}
\label{dvsint}
\delta_{v+s}^{interf.}(s,t)& = &\ap \left\{
(\qi\qf+\qis\qfs) \left[-\frac{1}{4}\lot{t^2}{s^2}\right)
-2 \mbox{Sp}(1+\frac{s}{t})+\frac{1}{2}\lo{t^2}{s^2}\right)\right]
\right. \nonumber\\
&-& \left. (\qi\qfs+\qis\qf) [t\rightarrow u]
-6-\frac{7}{6} \pi^2 \right\}.
\end{eqnarray}
Sp(z) denotes the Spence-function described in \cite{dilog}.
The integration over the scattering angle 
of the remnant of the IR-singular logarithm in Eq.~\ref{finterf} 
\begin{eqnarray}
\sigma^{(1)}_{interf.}(s)|_{\log.}& = & 
\int_{-1}^1 d\cos\!\theta
\; \frac{d\tilde{\sigma}^{(0)}(s,t)}{d\cos\!\theta}\;
\beta_{int.}(s,t)\; 
\lo{2 \Delta E}{\sqrt{s}}\frac{M_W^2}{|\dw-2 \sqrt{s} \Delta E|}\right)
\nonumber\\
&=& \tilde{\sigma}^{(0)}(s) \; \frac{\alpha}{\pi}\left(-\frac{1}{3}\right)
\; [5 (\qi\qf+\qis\qfs)+4(\qis\qf+\qfs\qi)] \times
\nonumber\\ 
& & \lo{2 \Delta E}{\sqrt{s}}\frac{M_W^2}{|\dw-2 \sqrt{s} \Delta E|}\right)
\end{eqnarray}
leads to a contribution, which will be completely compensated
by the hard photon contribution $\sigma_h^{interf.}(s)$ in
Eq.~\ref{hint} evaluated at $s=M_W^2$.
The remaining factor $\delta_{v+s}^{interf.}(s,t)$ together with the
IR-finite parts of the box diagrams 
$F_{V,(t,u)}^{finite}(s,t)$ (Eq.~\ref{intfin}),
where on-shell and mass singularities
have been subtracted according to Eq.~\ref{yfsmod},
are independent of the charge quantum numbers  
characterising the external fermions
\begin{equation}
\label{intrest}
(\delta_{v+s}^{interf.}+F_{V,(t,u)}^{finite})(s=M_W^2)
=-\ap\, \left[\dmw+8+\frac{5}{6}\, \pi^2\right] \; ,
\end{equation}
and, thus, can be absorbed into a modified weak contribution
to the differential Born-cross section.
This compensation of the non-factorisable 
$t(u)$-dependent remnants of the photonic box diagram
by $\delta_{v+s}^{interf.}$ is essential to
the factorisation of the numerator of the resonant cross section 
into partial $W$ widths describing the $W$ production and decay,
respectively. 

\subsection{The modified weak 1-loop correction to the $W$ production}

The IR-finite rest of the virtual photon contribution 
$F_{\gamma}^{finite}(s,t)$ of Eq.~\ref{photz} 
consists of the remnants of the YFS-prescription $F_{rem}^{\gamma}(s,t)$
and the IR-finite Feynman-diagrams III and IV
\begin{equation}
\label{finall}
F_{\gamma}^{finite}(s,t) = F_{rem.}^{\gamma}(s,t)
+(F_{I\!I\!I,f}^{\gamma}+F_{I\!I\!I,i}^{\gamma}
+F_{I\!V}^{\gamma})(s)|_{subtr.}
\end{equation}
with
\begin{equation}
\label{finite}
F_{rem.}^{\gamma}(s,t)  = \left\{\sum_{j=I,I\!I} (F_{j,f}^{finite}
+F_{j,i}^{finite})(s)
+(F_{V,t}^{finite}+F_{V,u}^{finite})(s,t)\right\}_{subtr.}\; ,
\end{equation}
where $|_{subtr.}$ reminds of the subtraction of the 
mass and on-shell singularities described by Eq.~\ref{yfsmod}.
After taking into account the remaining part of the
interference QED-form factor
$\delta_{v+s}^{interf.}(s,t)$ from Eq.~\ref{dvsint}, as  
it has already been discussed in Sec.~3.1,
these photon contributions can be absorbed 
into a modified weak contribution to the $W$ resonance
\begin{equation}
\tilde{F}_{weak}(s=M_W^2) = (F_{weak}^i+F_{weak}^f+
F_{\gamma}^{finite}+\delta_{v+s}^{interf.})(s=M_W^2)\; ,
\end{equation}
where $F_{weak}^{i,f}$ denote the pure weak contributions given
by Eq.~\ref{purewf}.
With this UV-finite and $\xi_i$-independent form factor the 
separation of the electroweak corrections to the $W$ resonance aimed for
is completed. 

Finally, it remains to check, whether $\tilde{F}_{weak}(M_W^2)$
can be represented as a sum of the modified weak corrections
to the $W$ width: $\delta\tilde{\Gamma}_{weak}^f$
and $\delta\tilde{\Gamma}_{weak}^{i}$.
According to Eq.~\ref{modwweak} this is equivalent to the 
verification of the identity
\[(F^{finite}_{\gamma}+\delta_{v+s}^{interf.})(s=M_W^2)
\equiv 2\; \delta\Gamma_{rem.}^{\gamma}\]
with $\delta\Gamma_{rem.}^{\gamma}$ given by Eq.~\ref{wremgam}.
In fact, by performing its explicit calculation   
this identity is proven to be true and
$\tilde{F}_{weak}(M_W^2)$ can be written as follows:
\begin{eqnarray}
\label{formweakm}
\tilde{F}_{weak}(M_W^2) 
& = &  \underbrace{F_{weak}^i(M_W^2)+\delta\Gamma_{rem.}^{\gamma}}_{=: \;
\tilde F_{weak}^{i}(M_W^2)}
+\underbrace{F_{weak}^f(M_W^2)
+ \delta\Gamma_{rem.}^{\gamma}}_{=: \;
\tilde{F}_{weak}^{f}(M_W^2)}
\nonumber\\
&\equiv & \delta\tilde{\Gamma}_{weak}^i+\delta\tilde{\Gamma}_{weak}^f\; .
\end{eqnarray}

By using this result and by following the prescription
derived in App.~A the $W$ production cross section
in the vicinity of the resonance including
(modified) weak 1-loop corrections has Breit-Wigner-form
\begin{equation}
\label{sigweak}
\sigma_w(s) = \frac{6 \pi}{M_W^2} \,\frac{(5-N_c^i)}{N_c^{i^2}}\, 
\frac{s\; \tilde{\Gamma}_{W\rightarrow f\!f'}^{(0+1)} \;  
\tilde{\Gamma}_{W\rightarrow i i'}^{(0+1)}}{[(s-M_W^2)^2+M_W^2 \; 
(\Gamma_W^{(0+1)})^2]} \; ,
\end{equation}
where $\tilde{\Gamma}$ denotes the QED-subtracted $W$ width
defined by Eq.~\ref{modwidth}. 

\section{Summary}
\setcounter{equation}{0}\setcounter{footnote}{0}

In order to match the requirements of future precision experiments at LEP II
and the Tevatron the corresponding cross sections for resonant
$W$ production have to be calculated beyond leading order 
perturbation theory. Having in mind the successful 
treatment of the electroweak ${\cal O}(\alpha)$ contribution
to the $Z$ resonance \cite{cern}, we strove for the analogous
description of the resonant $W$ production
in a 4-fermion process at the required level of
accuracy. After a thorough perturbative discussion
of the electroweak ${\cal O}(\alpha)$ contribution to 
the $W$ production we succeeded in extracting a
gauge invariant QED-like form factor from the photon contribution.
We showed, that, when approaching the $W$ resonance, the occurrence of 
on-shell singularities is
restricted to the initial state contribution
and can be 'regularised' by introducing the
$W$ width as a physical cut-off parameter in a gauge invariant way.
The similar structure of the resulting initial state QED-form factor 
to that of the $Z$ resonance allowed us to apply the same
technique to cope with the enhancement of the electroweak
coupling by large mass singular logarithms ({\em soft photon exponentiation}).
By separating the electroweak 1-loop corrections to 
the $W$ width into QED and weak contribution, too,
it turned out, that the (modified) weak corrections to
the resonant $W$ production cross section also factorises
into QED-subtracted partial $W$ widths.

In summary, we achieved a representation 
of the electroweak radiative corrections to the
W production cross section in the vicinity of the resonance,
which is in analogy to deep inelastic hadronic scattering 
a convolution of a process specific  
`hard' cross section $\sigma_w(s)$ (Eq.~\ref{sigweak})
with an universal radiator function
$G(z)$ (Eq.~\ref{radiator}) describing the initial state
photon contribution, where
the possibility of multiple soft photon emission
has been taken into account
\begin{equation}
\label{falt}
\fbox{$ \displaystyle \sigma(s) = \int_0^1 dz \; G(z)\; 
\sigma_w(sz) \; (1+\delta_{Q\!E\!D}^f) $} \; .
\end{equation}
$\delta_{Q\!E\!D}^f$, defined by
Eq.~\ref{qedfind}, denotes the final state photon
contribution, which is free of large mass singular logarithms.
As a result of the comparative discussion of the
S-matrix inspired ansatz and the perturbative
approach a transformation of the parameter of the resonance (Eq.~\ref{trafo})
connects between the two descriptions. 

\section{Numerical discussion}
\setcounter{equation}{0}\setcounter{footnote}{0}

In the following the numerical relevance of the 
different contributions to the electroweak radiative corrections
and their impact on the line shape of the $W$ resonance will be discussed.
For the numerical evaluation the following set of parameters 
has been used \cite{topfit},\cite{koba}:\\
\renewcommand{\arraystretch}{1}
\begin{table}[h]\centering
\begin{tabular}{ll}
$\alpha = 1/137.0359895$ &  $G_{\mu} = 1.16639 \cdot 10^{-5}
\; \mbox{GeV}^{-2}$ \\
$\alpha_s = 0.123$     &    $M_Z = 91.1884$ GeV \\ 
$m_d=m_u=0.0468$ GeV & $m_c=1.55$ GeV \\
$m_s=0.17$ GeV & $m_b = 4.7$ GeV \\
$|V_{ud}| = 0.975$ & $|V_{cs}| = 0.974$ \\
$|V_{tb}| = 0.999$ & $|V_{us}|=|V_{cd}| = 0.222$ \\
$|V_{cb}|=|V_{ts}|=0.044$ & $|V_{ub}| = |V_{td}|=0.007$ 
\end{tabular} 
\end{table}

\noindent
The masses of the light quarks are effective
quark masses in the sense, that they reproduce the correct
hadronic vacuum polarisation given by the dispersion integral 
calculated in \cite{jeg} and have no further physical meaning.
Using this set of input parameters
the $W$ boson mass is determined via the relation 
\begin{equation}
\label{mwt}
M_W^2 = \frac{M_Z^2}{2} \left[1+\sqrt{1-\frac{4 \pi \alpha}{\sqrt{2}G_{\mu}}
\frac{1}{M_Z^2}\frac{1}{1-\Delta r}}\right]
\end{equation}
as a function of the not precisely known or even unknown
parameters of the MSM: $m_t$ and $M_H$. 
A detailed description of $\Delta r$, which comprises the radiative
corrections to the muon decay, can be found in \cite{holl},\cite{lepewg}.

The $W$ width is an important ingredient of the description of 
the resonant $W$ boson production. The numerical results 
for the $W$ width at leading order $\overline{\Gamma}_W^{(0)}$
(Eq.~\ref{gammatot}) and at next-to-leading order 
$\overline{\Gamma}_W^{(0+1)}$ (Eq.~\ref{gammagm})
are summarised in Tab.~2. Besides the electroweak ${\cal O}(\alpha)$ 
contribution
calculated in App.~B, the latter contains also the contribution of virtual
and real gluons, so that $\overline{\Gamma}_W^{(0+1)}$ yields
in $G_{\mu}$-representation (Eq.~\ref{gmudar}) as follows:
\begin{equation}
\label{gammagm}
\overline{\Gamma}_W^{(0+1)}= \sum_{(ff'),f \not= t}
\overline{\Gamma}_{W\rightarrow f\!f'}^{(0)}
\; (1+2 {\cal R}\!e \, \delta \tilde{\Gamma}_{weak}^f-
\Delta r +\delta_{Q\!E\!D}^f
+\frac{N_c^f-1}{2}\; \delta_{Q\!C\!D}) \; ,
\end{equation}
where the modified weak correction and the QED-form factor are given by
Eq.~\ref{modwweak} and Eq.~\ref{qedfind}, respectively.
The QCD corrections are derived
in the limit of massless decay products \cite{lepwork}
\begin{equation}
\label{dqcd}
\delta_{Q\!C\!D} = \frac{\alpha_s}{\pi}\; \left[1+1.40932\;
 \left(\frac{\alpha_s}{\pi}\right)
-12.76706 \; \left(\frac{\alpha_s}{\pi}\right)^2\right] \; ,
\end{equation}
which for our case represents a sufficient approximation.
In the course of the calculation of the $W$ width the 
Kobayashi-Maskawa-mixing has been neglected, but
the final result has been multiplied with the square of the
corresponding physical matrix element $V_{ij}$.
From a numerical point of view, this
procedure does not significantly differ 
from a consideration of the Kobayashi-Maskawa-matrix 
in the renormalisation procedure as it has been pointed out in \cite{ths}.
In order to illustrate the variation of $M_W$ and 
$\overline{\Gamma}_W^{(0+1)}$ with the electroweak input parameters, 
they are given in Tab.~2 for different values of $m_t$ and $M_H$.
The ratio $\overline{\Gamma}_W^{(0+1)}/M_W$ illustrates the very weak 
dependence of the $W$ width
on $m_t$ and $M_H$: due to the cancellation of large 
leading (quadratic) $m_t$-dependent contributions in
$\delta \tilde{\Gamma}_{weak}$ and $\Delta r$
only a logarithmic dependence on $m_t$ (and $M_H$) survives
and thus the variation of
$\overline{\Gamma}_W^{(0+1)}$ is mainly a consequence of
the variation of $M_W$. Our result obtained for the 
$W$ width in next-to-leading order is in 
very good agreement with the total $W$ width derived in \cite{width}:
relative deviation $\le 0.005 \%$. \\
\begin{table}[htb]\centering
\label{gamw}
\begin{tabular}{lrrr}
\hline\\
$M_H$ [GeV] & $60$  & $300$ & $1000$ \\ \hline
$m_t=165$ GeV& & & \\ \hline
$M_W$ [GeV]                        & 80.3648 & 80.2618 & 80.1647 \\
$\overline{\Gamma}_W^{(0)}$ [GeV]  &  2.0433 &  2.0354 &  2.0280 \\
$\overline{\Gamma}_W^{(0+1)}$ [GeV]&  2.0911 &  2.0834 &  2.0759 \\ 
$\overline{\Gamma}_W^{(0+1)}/M_W$  &  0.0260 &  0.0260 &  0.0259 \\ \hline
$m_t=175$ GeV& & & \\ \hline
      & 80.4275 &  80.3228   & 80.2244 \\   
      &  2.0481 &   2.0401   &  2.0326  \\ 
      &  2.0960 &   2.0882   &  2.0806 \\
      &  0.0261 &   0.0260   &  0.0259  \\  
\hline
$m_t=185$ GeV& & &\\\hline
& 80.4927 &  80.3861 & 80.2862   \\
&  2.0531 &   2.0449 &  2.0373 \\ 
&  2.1012 &   2.0932 &  2.0854 \\
&  0.0261 &   0.0260 &  0.0260   \\  
\hline
\end{tabular}
\caption{The total $W$ width (and $M_W$) in $G_{\mu}$-representation
including the described radiative corrections}
\end{table}

In the subsequent discussion of the line shape of the $W$ resonance 
the top quark mass 
and the $W$ boson mass are chosen to be the central values of
their current world average (\cite{topm} and \cite{mwav}, resp.)
\[ m_t=175 \pm 9 \; \mbox{GeV}  \]
\[ M_W=80.33 \pm 0.15\; \mbox{GeV} \; .\]
Using these input parameters the Higgs-boson mass and the 
total $W$ width yield
\[M_H = 273 \;  \mbox{GeV}\;  \Rightarrow \; \overline{\Gamma}_W^{(0)} = 2.0406
\; \mbox{GeV} \; \; \;  \mbox{and} \; \; \;
\overline{\Gamma}_W^{(0+1)} = 2.0887\; \mbox{GeV} \] 
compared to the measured value of $\Gamma_W$ \cite{koba}
\[\Gamma_W = 2.08 \pm 0.07 \; \mbox{GeV} \; .\] \\

\underline{{\bf The `hard' cross section $\sigma_w(s)$}} \\

The effect of the
(modified) weak 1-loop correction described by Eq.~\ref{sigweak} to the
$W$ line shape is shown in Fig.~4 for the example of 
a pure leptonic process: $\nu_e e^+\rightarrow \nu_{\mu} \mu^+$.
There is no noticeable impact on 
the location of the maximum of the resonant cross section
$\overline{\sigma}_w(s)$ (in $G_{\mu}$-representation) 
\begin{equation}
\label{peakp}
s_{max}=M_W^2\; \sqrt{1+\gamma^2} \; ,
\end{equation}
where the abbreviation
$\gamma=\frac{\textstyle{\overline{\Gamma}_W^{(0+1)}}}{\textstyle{M_W}}$
has been used, due to the smallness of $\gamma$ in the above equation
($\Delta s_{max}=0.6$ MeV).
The maximum of the cross section, however,
\begin{equation}
\label{wmax}
\overline{\sigma}_{w,max} = \frac{6 \pi}{M_W^2} \; 
\frac{5-N_c^i}{N_c^{i^2}} \;  
\frac{\tilde{\overline{\Gamma}}_{W\rightarrow f f'}^{(0+1)} \;  
\tilde{\overline{\Gamma}}_{W\rightarrow i i'}^{(0+1)}}
{(\overline{\Gamma}_W^{(0+1)})^2}
\; (1+\frac{1}{4} \gamma^2) 
\end{equation}
is reduced as compared to the peak value in leading order perturbation theory
$\overline\sigma^{(0)}_{max}$ ($=\overline\sigma_{w,max}$ with
$\Gamma^{(0+1)} \rightarrow \Gamma^{(0)}$).
For the case of the leptonic process this reduction yields
\[ \overline{\sigma}_{w,max} = 0.9347 \; \overline\sigma^{(0)}_{max} \]
and is mainly due to the QCD correction to the total $W$ width 
given by Eq.~\ref{dqcd}. Thus, when considering the
$W$ production process
$\nu_e e^+ \rightarrow u \overline d$ the reduction of the maximum
cross section only amounts to
\[ \overline{\sigma}_{w,max} = 0.9729 \; 
\overline\sigma^{(0)}_{max} \; ,\]
since now the QED-subtracted partial $W$ width
$\tilde{\overline{\Gamma}}_{W\rightarrow u\overline d}^{(0+1)}$
of Eq.~\ref{wmax} also includes the QCD contribution.
Tab.~3 shows the negligible small dependence of the peak value 
$\overline{\sigma}_{w,max}$ on the 
top quark and Higgs-boson mass due to the aforementioned
cancellation
of leading (quadratic) $m_t$-dependent contributions in the 
partial $W$ width calculated in the $G_{\mu}$-representation.\\
\renewcommand{\arraystretch}{1.5}
\begin{table}[htb]\centering
\label{gamwb}
\begin{tabular}{|c||r|c|c|} \hline 
$m_t\,$ [GeV] & $M_H\,$  [GeV]& $\overline{\Gamma}_W^{(0+1)}\,$ [GeV] &
$\overline{\sigma}_{w,max}\, $[nb] \\ \hline \hline
166  & 124.19  & 2.0886 & 52.5449 \\ \hline
175  & 273.32  & 2.0887 & 52.5451 \\ \hline
184  & 549.30  & 2.0888 & 52.5452 \\ \hline
\end{tabular}
\caption{The $W$ width $\overline{\Gamma}_W^{(0+1)}$ and
the peak value $\overline{\sigma}_{w,max}$ for different top quark masses.
Besides the top quark mass the $W$ boson mass $M_W = 80.33$ GeV
has been used as an input parameter,
so that the Higgs-boson mass is determined by Eq.~5.1.}
\end{table}

The further discussion is dedicated to the QED-like contribution,
especially to the initial state photon radiation.
The final state QED contribution described by  
$\delta_{Q\!E\!D}^f$ of Eq.~\ref{qedfind}
has a tiny effect on the peak value:
$\delta_{Q\!E\!D}^{f=\mu}\sim 0.0072$ for leptons
and $\delta_{Q\!E\!D}^{f=u} \sim 0.0069$ for quarks, but has no impact 
on the peak position of the resonant cross section.
The leftovers of the interference contribution have already been 
absorbed in the `hard' cross section as it has been described in Sec.~3.1.\\

\underline{{\bf The {\em initial state} bremsstrahlung}}\\

The initial state bremsstrahlung, described by 
Eq.~\ref{sigsin} (soft photons) together with 
Eq.~\ref{hin} (hard photons), does not only carry the 
main contribution to the reduction of the peak value, but is
also responsible for the distortion of the line shape, especially for the 
shift in the peak position. The main effect to the
reduction of the maximum can roughly be estimated by the factor
\[ 1-\beta_{i=e}(M_W^2) \; \lo{M_W}{\overline{\Gamma}_W^{(0+1)}}\right)
 = 0.81\]
with $\beta_{i=e}(M_W^2)$ given by Eq.~\ref{beti}.
For comparison, the corresponding factor for the case of the $Z$ resonance
is given by \cite{holl}
\[ 1-4\frac{\alpha}{\pi}\lo{M_Z}{m_e}\right)\lo{M_Z}{\Gamma_Z}\right)
=0.6 \; .\]
The effect is much smaller, when the soft photon is emitted by quarks
\[ 1-\beta_{i=u}(M_W^2) \; \lo{M_W}{\overline{\Gamma}_W^{(0+1)}}\right)
 = 0.94 \; ,\]
where the numerical evaluation has been performed by using the 
effective quark masses.
They have no physical meaning, but in a realistic hadronic scattering process
they are rather included in the
parton distribution as parts of the interacting hadrons,
with which the parton cross section has to be convoluted 
in order to obtain an observable cross section \cite{kripf}.
 
In Fig.~5 the impact of the initial state bremsstrahlung 
to the $W$ line shape in a pure leptonic process
$\nu_e e^+ \rightarrow \nu_{\mu} \mu^+$ is shown.
The shift of the peak position due to the energy loss in the resonant
$W$ propagator in ${\cal O}(\alpha)$ amounts to
\[  \Delta M_W = +53 \; \mbox{MeV} \; ,  \]
which reduces to 
\[\Delta M_W=+42 \; \mbox{MeV} \] 
after performing soft photon exponentiation as it is described
by Eq.~\ref{falti}.
This shows, that the calculation performed
in ${\cal O}(\alpha)$ overestimates the $W$ boson mass by
11 MeV. Due to the different charge structure for the case of 
quarks in the initial state only a shift
of the peak position by $\Delta M_W=+14$ MeV 
can be observed, which still amounts to $\Delta M_W=+13$ MeV
after the resummation of the soft photon contribution.
Since these soft photons represent the main contribution 
to the resonant $W$ production, we expect no significant
contribution from hard photons at 2-loop level,
which has been confirmed by an explicit 2-loop calculation 
in the case of the $Z$ resonance \cite{bcern}.

In summary, the electroweak ${\cal O}(\alpha)$ contribution to
the resonant $W$ production develops the same characteristics
as the corresponding corrections to the $Z$ line shape.
Fig.~6 shows the total cross section of the $W$ production
in the vicinity of the resonance as it is 
described by the convolution integral of Eq.~\ref{falt}, where
the $s$-dependence of the $W$ width has been considered by
applying the transformations of Eq.~\ref{trafo}. 
The main impact of the discussed
radiative corrections on the $W$ line shape can be summarised as
follows:
\begin{itemize}
\item
The peak position $s_{max}$ of the resonant cross section
(Eq.~\ref{peakp})
is shifted about $+42$ MeV ($Z:+96$ MeV) (constant $W$ width)
and suffers an additional shift about $-27$ MeV ($Z:-34$ MeV), when
assuming an $s$-dependent width.
\item
The peak value of the resonant cross section
is reduced by a factor 0.82 ($Z: \sim$0.6) with respect to 
$\overline{\sigma}^{(0)}_{max}$.
\end{itemize}
For comparison, the corresponding values in case of the $Z$ resonance
are also provided \cite{holl} (in brackets).

\section*{Acknowledgments}

The Fermi National Accelerator Laboratory is operated by Universities Research
Association, Inc.~, under contract DE-AC02-76CHO3000 with the
U.S.~Department of Energy.

\begin{figure}
\vspace{9cm}
\includegraphics{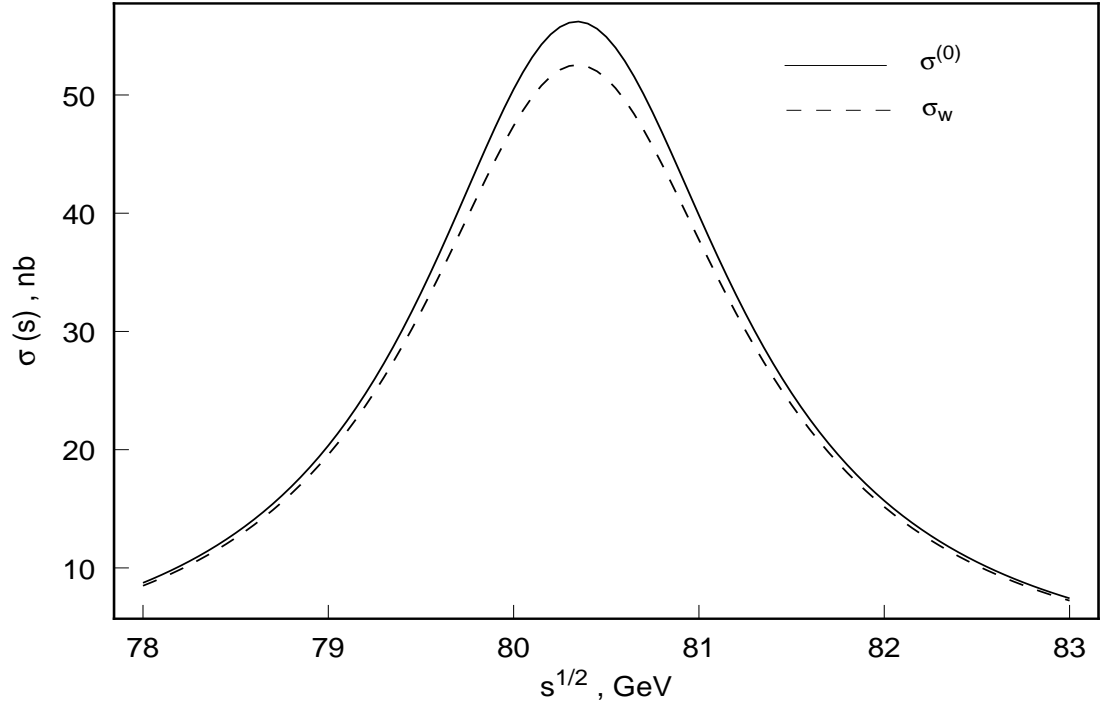}
\caption{The `hard' cross section $\overline\sigma_w(s)$ of Eq.~3.45
compared to the Born-cross section
for $\nu_e e^+\rightarrow \nu_{\mu} \mu^+$}
\end{figure}

\begin{figure}
\vspace{8cm}
\includegraphics{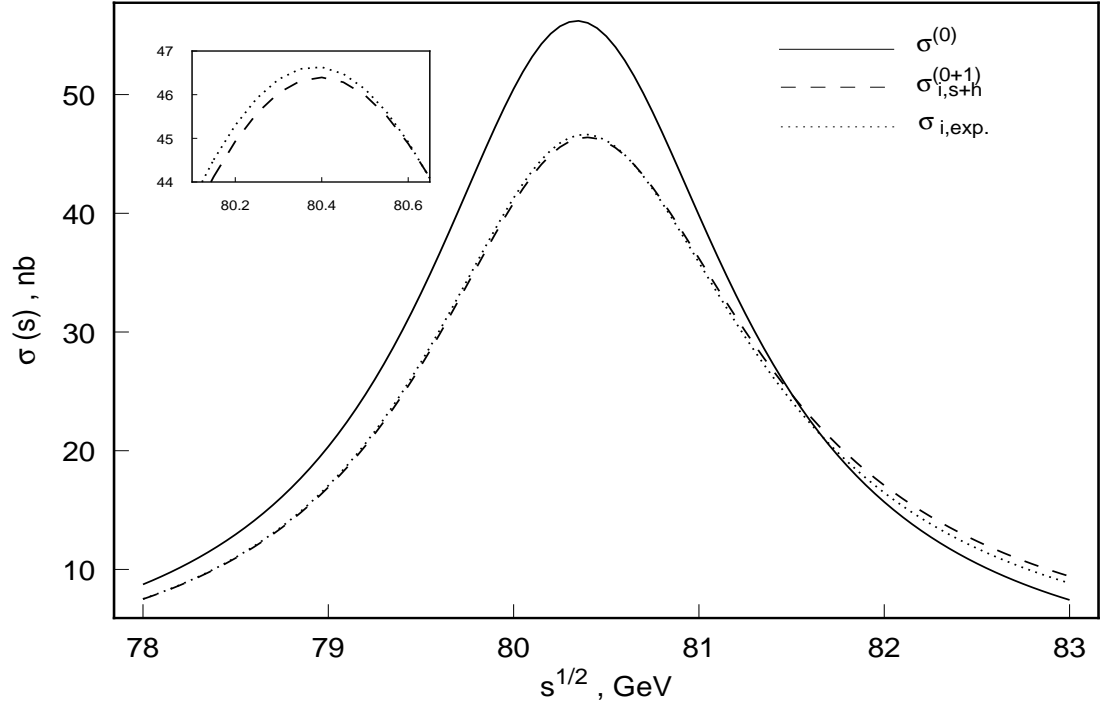}
\caption{The effect of initial state bremsstrahlung in ${\cal O}(\alpha)$
described by $\sigma_{i,s+h}^{(0+1)}(s)$ of Eq.~3.29
and after soft photon exponentiation (Eq.~3.31)}
\end{figure}

\begin{figure}
\vspace{8cm}
\includegraphics{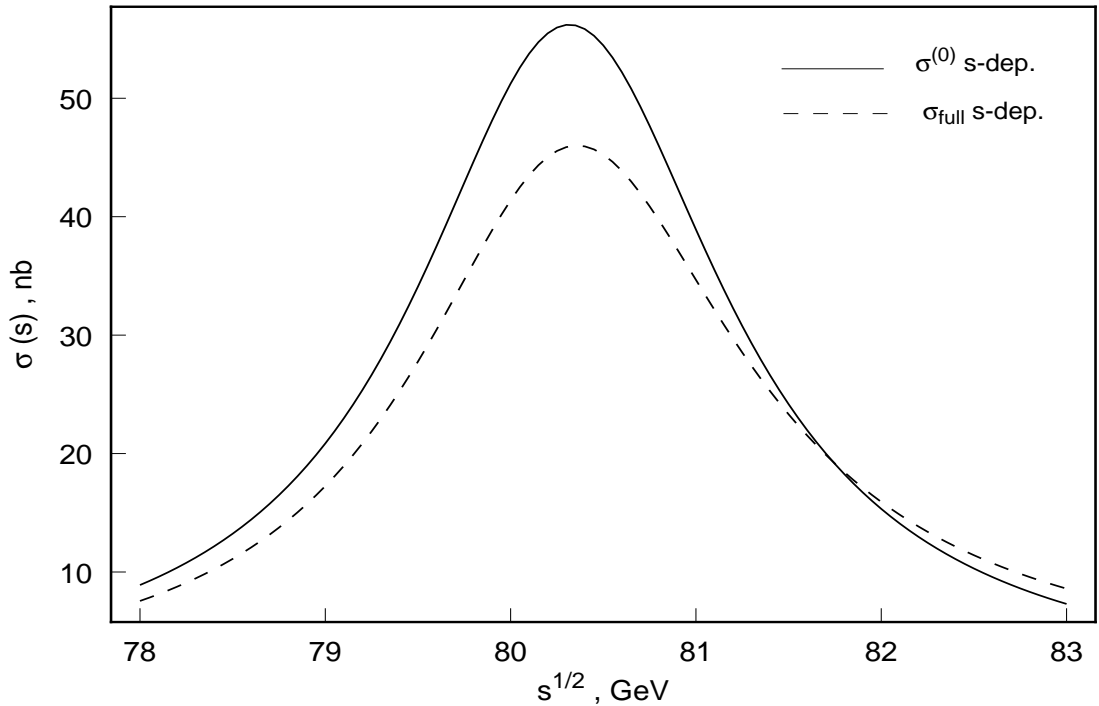}
\caption{The $W$ production cross section in the vicinity of the
resonance including the discussed electroweak radiative corrections (Eq.~4.1)}
\end{figure}

\newpage

\noindent
{\huge \bf Appendix}

\begin{appendix}

\section{Unstable particles and gauge invariance}
\setcounter{equation}{0}\setcounter{footnote}{0}

In S-matrix theory an unstable particle, experimentally seen as 
a resonance during the interaction of stable particles, can be
easily described when neglecting all singularities besides a
single complex pole close to the real energy axes with negative
imaginary part \cite{polk}.
Therefore the S-matrix is approximately of the form of a 
Breit-Wigner resonance
\begin{equation}
\label{smat}
{\cal M}(s) = \frac{R}{s-M_c^2}+F(s) \; , 
\end{equation}
where $F(s)$ is an analytic function with no poles.
The residue $R$ of the complex pole $M_c^2$ can be interpreted as
a product of coupling constants, with which the unstable particle 
couples to the external particles \cite{polk}. The resonance in the 
scattering amplitude arises in the vicinity of 
$s={\cal R}\!e(M_c^2)$, the physical mass of the unstable particle, and the
width of the resonance is given by ${\cal I}\!m(M_c^2)$:
\begin{equation}
\label{mphys}
M_c^2 = M_{phys.}^2-i M_{phys.} \Gamma \; \; \mbox{    or e.g.  }\; \;
M_c^2 = (M_{phys.}-i \frac{\Gamma}{2})^2 \; .
\end{equation} 
The S-Matrix given by Eq.~\ref{smat} is gauge invariant in the physical region 
($s$ real and $s>0$)
and thus -via analytic continuation- also in the complex energy plane,
which enables its application in a gauge theory.
The fact that the complex pole $M_c$, its residue $R$ 
and the non-resonant part $F(s)$
are separately gauge invariant has been used to find a 
gauge invariant description of the $Z$ resonance at the required level
of accuracy \cite{zres}.

From the quantum field theory's point of view
a resonance in the scattering amplitude
is caused by a pole in the propagator of an unstable particle. 
In the vicinity of the resonance the resummed propagator has to be used,
which is a formal summation of a geometric series with the 
1PI self energy of the unstable particle as argument (Dyson-resummation)
\cite{dys}:\\

\hspace{1cm} \psfig{figure=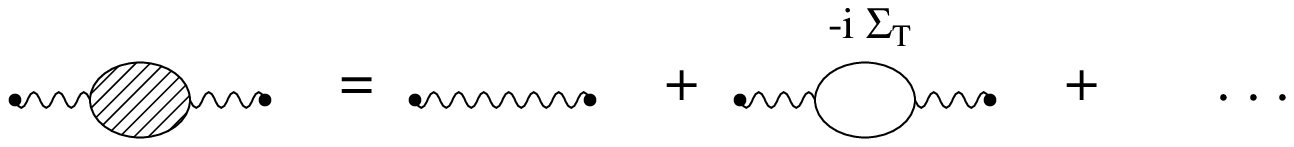}

\begin{equation}
\label{prop}
D^{\mu \nu} = 
\frac{-i g^{\mu \nu}}{s-M_0^2+\Sigma_T(s)} 
= \frac{-i g^{\mu \nu}}{s-M_0^2+i\eps} \;
\left[1+\left(\frac{-\Sigma_T(s)}{s-M_0^2+i\eps}\right)+\ldots\right] \; .
\end{equation}
$M_0$ denotes the unrenormalised bare mass and $\Sigma_T(s)$ is the 
transverse part of the 1PI self energy. Since
the external particles are considered to be massless as long as no
singularities occur, the longitudinal part of the propagator does not
contribute and will not be discussed.
Veltman \cite{veltb} showed that the S-matrix constructed by using
the Dyson-resummed propagator and 
assuming only transitions between stable particles obeys the
principles of unitarity and causality. 
Thus, the field theoretical description of 
gauge boson resonances is given by the following amplitude,
after performing a renormalisation procedure:
\begin{equation}
\label{pert}
{\cal M}(s) = \frac{\hat V_i(s)\; \hat V_f(s)}{s-
M_R^2+\hat\Sigma_T(s)}+B(s) \; .
\end{equation}
$\hat V_{i,f}(s)$ denote the renormalised vertices, describing the
production and decay of the unstable particle, $M_R$
denotes the renormalised mass and $\hat \Sigma_T(s)$
the renormalised self energy. $B(s)$ comprises the non-resonant contributions,
e.g. box diagrams.

The S-matrix theory inspired construction of a gauge invariant amplitude
using a Laurent expansion of Eq.~\ref{pert} around the complex pole and 
afterwards performing a consistent evaluation of the parameters
of the resonance in the coupling constant $g$ results in a description
with constant width.  
Choosing the field theoretical ansatz and carrying out
a consistent treatment of the inverse of the propagator in Eq.~\ref{pert}
can lead to a scattering amplitude with $s$-dependent width \cite{wetzel}.
Analysing the $Z$ line shape in the S-matrix theory approach yields a  
$Z$ boson mass which is about 34 MeV larger, at ${\cal O}(g^2)$ accuracy,
than the corresponding value obtained in an $s$-dependent width prescription.
Since these two descriptions are 
connected by a transformation of the line shape parameters
\cite{bard} there are equivalent and, thus,
the difference in the $Z$ boson mass has no physical meaning.

The future precise measurement of the $W$ boson mass at LEP II and at 
an upgrade of the Tevatron raises the same questions 
for a charged gauge boson resonance. 
In the following, the applicability of the prescriptions, derived
in the context of the $Z$ resonance, to a charged vector boson resonance
will be studied.\\

\underline{{\bf ${\cal M}_{ii'\rightarrow ff'}(s)$ with constant width}}\\

Following the treatment given in \cite{zres},
which can be directly applied to the $W$ resonance, a gauge invariant 
scattering amplitude and a definition of mass and width 
to the required level of accuracy can be given:
\begin{itemize}
\item \underline{${\cal O}(g^0)$ accuracy}:\\
At 1-loop level the physical mass $M_W$ is connected to
the renormalised mass as follows:
\begin{equation}
\label{massdef}
{\cal R}\!e(M_c^2)=M_W^2 = M_R^2 -{\cal R}\!e \hat\Sigma_T(M_R^2,g^2) \; ,
\end{equation}
which yields the equality of physical and renormalised mass when
using the on-shell renormalisation condition
${\cal R}\!e \hat\Sigma_T(M_R^2,g^2)=0$
in order to determine the mass renormalisation constant $\delta M_W^2
\equiv M_0^2-M_R^2$.
In leading order perturbation theory the 
$W$ width corresponds to the imaginary part of the 1-loop corrected 
renormalised $W$ self energy 
\begin{equation}
\label{gama}
M_W \; \Gamma_W^{(0)} = {\cal I}\!m \hat\Sigma_T(M_W^2,g^2) \; .
\end{equation}
Thus, the $W$ resonance is described by 
\begin{equation}
{\cal M}^{(0)}(s) = \frac{{\cal R}(g^2)}{s-M_W^2+i M_W \; \Gamma_W^{(0)}}
+ {\cal O}(g^2)
\end{equation}
with
\[{\cal R}(g^2)  =  V_i(g) V_f(g) \; .\]
\item \underline{${\cal O}(g^2)$ accuracy}:\\
In next-to-leading order Eq.~\ref{massdef} turns to
\begin{eqnarray}
M_W^2 & = & M_R^2 - (1-{\cal R}\!e \hat\Pi_T(M_R^2,g^2))\;
{\cal R}\!e \hat\Sigma_T(M_R^2,g^2) \nonumber \\
& - & {\cal R}\!e \hat\Sigma_T(M_R^2,g^4)
- {\cal I}\!m \hat\Sigma_T(M_R^2,g^2) \; {\cal I}\!m \hat\Pi_T(M_R^2,g^2) \; ,
\end{eqnarray}
where the following abbreviation has been used:
\[\hat\Pi_T(s) \equiv \frac{\partial \hat \Sigma_T(s)}{\partial s} \; .\]
Taking the renormalised mass as the zero of the real part 
of the inverse propagator, which corresponds to the
field theoretical definition of a stable particle's mass,
this reduces to
\begin{equation}
M_W^2 = M_R^2 - {\cal I}\!m \hat\Sigma_T(M_R^2,g^2)
\; {\cal I}\!m \hat\Pi_T(M_R^2,g^2) \; .
\end{equation}
Thus, one obtains a shifted renormalised mass with respect to the 
physical mass. By considering a renormalisation condition, however,
which reads at 2-loop level as follows:
\begin{equation}
\label{ren}
{\cal R}\!e \hat\Sigma_T(M_R^2,g^4)+
{\cal I}\!m \hat\Sigma_T(M_R^2,g^2) \; {\cal I}\!m \hat\Pi_T(M_R^2,g^2)
 = 0 \; ,
\end{equation}   
the equality of physical and renormalised mass is recovered \cite{zres}.
Then the $W$ width in next-to-leading order yields  
\begin{equation}
\label{ww}
M_W \Gamma_W^{(0+1)} =  (1-{\cal R}\!e \hat\Pi_T(M_W^2,g^2))\;
{\cal I}\!m \hat\Sigma_T(M_W^2,g^2)+{\cal I}\!m \hat\Sigma_T(M_W^2,g^4) \; .
\end{equation}
The calculation of $\Gamma_W^{(0+1)}$ in the MSM and for $\xi_i=1$
can be found in \cite{width} and will be additionally performed
in App.~B for $R_{\xi}$-gauge and in the limit of massless
decay products.
Finally, a gauge invariant description of the $W$ resonance can be given,
which completely takes into account the electroweak radiative corrections
up to order ${\cal O}(g^2)$
\begin{equation}
\label{mat}
{\cal M}^{(0+1)}(s)  = \frac{{\cal R}(g^2) + {\cal R}(M_W^2,g^4)}
{s-M_W^2+i M_W \; \Gamma_W^{(0+1)}}+ {\cal O}(g^4)
\end{equation}
with the residue in next-to-leading order
\begin{equation}
\label{rescon}
{\cal R}(M_W^2,g^4) = \hat V_i(M_W^2,g^3) V_f(g)
+V_i(g) \hat V_f(M_W^2,g^3)-V_i(g) V_f(g) \; \hat\Pi_T(M_W^2,g^2)\; . 
\end{equation}
The $\hat V_{i,f}(M_W^2,g^3)$ denote the renormalised vertices including
1-loop corrections to the production and decay of a $W$ boson, respectively.
\end{itemize}

\underline{{\bf ${\cal M}_{ii'\rightarrow ff'}(s)$ with $s$-dependent width}}\\

Next we present the results obtained by using the field theoretical ansatz
and we discuss the equivalency 
of both approaches also for a charged gauge boson resonance.
The latter cannot be readily expected in the case of a
$W$ resonance, since the existence of a transformation given by Bardin 
et al.~\cite{bard} for the case of the $Z$ resonance
is due to the linear $s$-dependence of the imaginary part of
the $Z$ self energy. Therefore a careful study of the $s$-dependence of the 
$W$ self energy is required.
After evaluating the real part of the $W$ self energy in Eq.~\ref{prop} 
(after renormalisation)
around $s=M_R^2$ and using the on-shell renormalisation condition
${\cal R}\!e \hat \Sigma_T(M_R^2)=0$ the $W$ propagator is given by
\begin{equation}
\label{propw}
D_W^{\mu\nu}= -ig^{\mu \nu} \; \frac{1-{\cal R}\!e \hat\Pi_T(M_R^2)}
{s-M_R^2+i {\cal I}\!m \hat\Sigma_T(s)\; (1-{\cal R}\!e \hat\Pi_T(M_R^2))} \; .
\end{equation}
Thus, following the argumentation of Wetzel \cite{wetzel}
in the vicinity of the resonance
the residue of the complex pole in Eq.~\ref{pert}
in next-to-leading order is given by
\begin{equation}
R^{(0+1)}(M_W^2) = \hat V_i(M_W^2,g^3) V_f(g)
+V_i(g) \hat V_f(M_W^2,g^3)+V_i(g) V_f(g) \; (1-
{\cal R}\!e \hat\Pi_T(M_W^2,g^2)) \; ,
\end{equation}
where $M_R = M_W$ has been used.
Since the inverse $W$ propagator is of order $g^2$
in the vicinity of the resonance
the complete ${\cal O}(g^4)$-contribution to the denominator has to be taken
into account. Thus, after using the definition of the $W$ width given 
by Eq. \ref{ww}, the following definition for the $s$-dependent $W$ width
can be given:
\begin{eqnarray}
\mbox{denominator} & = &
s-M_W^2+i M_W \Gamma_W^{(0+1)}+i {\cal I}\!m
[\hat\Sigma_T(s,g^2)-\hat\Sigma_T(M_W^2,g^2)]  
\nonumber\\
& =: & s-M_W^2+i M_W \Gamma_W^{(0+1)}(s) \; .
\end{eqnarray}
In contrary to the $Z$ boson, where the imaginary part of the
derivative of the 
1PI $Z$ self energy develops gauge dependent contributions only when 
\cite{sir}
\[\xi_W \leq\left(\frac{M_Z}{2M_W}\right)^2 \; ,\] 
the corresponding quantity in the $W$ boson case    
${\cal I}\!m \hat\Pi_T(M_W^2,g^2)$ is gauge parameter dependent for each
gauge parameter $\xi_i\neq 1$ (Eq.~\ref{imagself}).
This is due to the existence of Feynman-diagrams involving photons, which
couple to the $W$ boson via the triple gauge boson coupling.
The 1-loop contributions
to the $W$ self energy are shown in Fig.~7 for $R_{\xi}$-gauge.
\begin{figure}
\label{phot}
\hspace{0.5cm} \psfig{figure=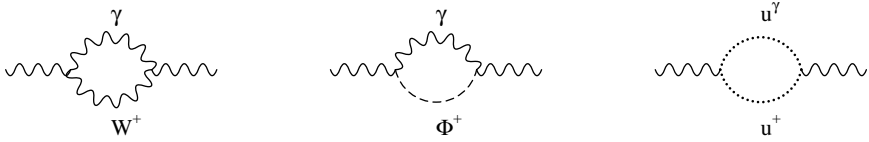}
\caption{Feynman-diagrams for the photonic 1-loop correction to the
$W$ self energy
(the dashed and dotted lines denote a charged Higgs-ghost
$\Phi^{\pm}$ and the Faddeev-Popov-ghosts $u^{\pm}$ or $u^{\gamma}$, resp.)}
\end{figure}
However, when the Dyson-resummed contribution
\[{\cal I}\!m [\hat\Sigma_T(s,g^2)-\hat\Sigma_T(M_W^2,g^2)] 
\stackrel{s\rightarrow M_W^2} =
(s-M_W^2) \; {\cal I}\!m \hat\Pi_T(M_W^2,g^2) \]     
is treated perturbatively in order to cancel the 
gauge parameter dependent contributions to the imaginary part of the 1PI
vertex corrections in $R^{(0+1)}(M_W^2)$,
the Breit-Wigner resonance formula with constant width from Eq.~\ref{mat}
in combination with the renormalisation condition of order
${\cal O}(g^4)$ given by Eq.~\ref{ren} is recovered.

In order to obtain the physical description of the $W$ resonance
with $s$-dependent width, the following approximation of the $s$-dependence
of the photon contribution to the
imaginary part of the $W$ self energy shown in Fig.~7
is useful ($I(s)$: Eq.~\ref{imag}):
\begin{eqnarray}
{\cal I}\!m \hat\Sigma_T^{\gamma}(s) & = & (s-M_W^2) \; 
\theta(s-M_W^2) \; I(s) \nonumber\\
& \stackrel{R} = & (s-M_W^2) \; \theta(s-M_W^2) \; I(M_W^2) 
:= (s-M_W^2) \; {\cal I}\!m \hat\Pi_T^{\gamma}(M_W^2) \; .
\end{eqnarray}
Since the derivative ${\cal I}\!m \hat\Pi_T^{\gamma}(M_W^2)$
does not exist in a strict mathematical sense
due to the threshold at $s=M_W^2$,
the above equation has to be understood as a definition.
The fermion contribution to ${\cal I}\!m \hat\Sigma_T(s)$, however, is 
linear in $s$ in the case of massless fermions, so that the
$s$-dependence can be extracted as follows:
\begin{equation}
\label{ima}
{\cal I}\!m \hat\Sigma_T(s,g^2) = \frac{s}{M_W^2} \; 
{\cal I}\!m \hat\Sigma_T(M_W^2,g^2)+(s-M_W^2)
\; {\cal I}\!m \hat\Pi_T^{\gamma}(M_W^2,g^2) \; .
\end{equation}
By using this $s$-dependence in the $W$ propagator
given by Eq.~\ref{propw} and after undoing the resummation of 
${\cal I}\!m \hat\Pi_T^{\gamma}(M_W^2,g^2)$ the $W$ propagator 
turns out to be as follows:
\begin{equation}
D_W^{\mu\nu}= -ig^{\mu \nu} \; \frac{1-{\cal R}\!e \hat\Pi_T(M_W^2,g^2)-
i {\cal I}\!m \hat\Pi_T^{\gamma}(M_W^2,g^2)+{\cal O}(g^4)}
{s-M_W^2+i \frac{s}{M_W^2}{\cal I}\!m \hat\Sigma_T(M_W^2) 
\; (1-{\cal R}\!e \hat\Pi_T(M_W^2,g^2)
-i {\cal I}\!m \hat\Pi_T^{\gamma}(M_W^2,g^2))+{\cal O}(g^6)} \; ,
\end{equation}
where the validity of Eq.~\ref{ima}
at least up to order $g^4$ has been assumed.
In summary, the scattering amplitude 
constructed with the help of this propagator and a subsequent
consistent evaluation in the coupling constant of the
numerator and the denominator, which results in a gauge invariant 
description of a resonant produced $W$ boson 
at the required level of accuracy, will be given: 
\begin{itemize}
\item \underline{${\cal O}(g^0)$ accuracy}:
\begin{equation}
{\cal M}^{(0)}(s) = \frac{R^{(0)}}{s-M_W^2+i \frac{s}{M_W}
\; \Gamma_W^{(0)}}+{\cal O}(g^2)
\end{equation}
with
\[R^{(0)} = V_i(g) V_f(g) \]
and the definition of the $W$ width given by Eq.~\ref{gama}.
\item \underline{${\cal O}(g^2)$ accuracy}: \\
By considering the following renormalisation condition: 
\begin{equation}
{\cal R}\!e \hat\Sigma_T(M_W^2,g^4)
+{\cal I}\!m \hat\Sigma_T(M_W^2,g^2) \; 
{\cal I}\!m \hat\Pi_T^{\gamma}(M_W^2,g^2) = 0 \; ,
\end{equation}
which differs from Eq.~\ref{ren} by 
\begin{equation}
\label{dif} 
M_W \Gamma_W^{(0)} \; {\cal I}\!m \hat\Pi_T^{ferm.}(M_W^2,g^2)=
(\Gamma_W^{(0)})^2
\end{equation}
the scattering amplitude is given by:
\begin{equation}
\label{matp}
{\cal M}^{(0+1)}(s)  = \frac{R^{(0+1)}(M_W^2)}
{s-M_W^2+i \frac{s}{M_W} \; \Gamma_W^{(0+1)}}+ {\cal O}(g^4)
\end{equation}
with
\begin{eqnarray}
R^{(0+1)}(M_W^2)& = & V_i(g) V_f(g)+\hat V_i(M_W^2,g^3) V_f(g)
+V_i(g) \hat V_f(M_W^2,g^3)
\nonumber\\
& - & V_i(g) V_f(g) \; [{\cal R}\!e 
\hat\Pi_T(M_W^2,g^2)+i {\cal I}\!m \hat\Pi_T^{\gamma}(M_W^2,g^2)] \; .
\end{eqnarray}
The next-to-leading order $W$ width
$\Gamma_W^{(0+1)}$ is again defined by Eq.~\ref{ww}.
$R^{(0+1)}(M_W^2)$ differs from ${\cal R}(M_W^2,g^4)$ of
Eq.~\ref{rescon} concerning their imaginary parts by
\begin{equation}
\label{diff}
V_i(g) V_f(g) \;
{\cal I}\!m \hat\Pi_T^{ferm.}(M_W^2,g^2)=
V_i(g) V_f(g) \;\frac{\Gamma_W^{(0)}}{M_W} \; .
\end{equation}
\end{itemize}

It remains to check whether both descriptions are equivalent.
For that purpose Eq.~\ref{matp} will be rewritten
as follows (with $\gamma = \frac{\Gamma_W^{(0+1)}}{M_W}$):
\begin{eqnarray}
{\cal M}^{(0+1)}(s) & = & \frac{R^{(0+1)}(M_W^2)}
{s \; (1+i \gamma)-M_W^2}
\nonumber\\
& = & \frac{R^{(0+1)}(M_W^2) \;\frac{(1-i \gamma)}{(1+\gamma^2)}}
{s-M_W^2\;(1-\gamma^2)+i M_W^2 (1-\gamma^2) \gamma}
\nonumber\\
& =: & \frac{\overline{R}^{(0+1)}(M_W^2)}
{s-\overline{M}_W^2+i \overline{M}_W\; \overline{\Gamma}_W^{(0+1)}}\; .
\end{eqnarray}
The evaluation of the   
numerator and denominator of the above equation up to the
order required for a ${\cal O}(g^2)$ accuracy easily verifies,
that exactly those terms arise, in which the
$s$-dependent width description differs from the 
constant width amplitude given by the Eqs.~\ref{dif}, \ref{diff}.
Thus, a transformation of the parameters of the resonance:
residue, position of the pole ($\rightarrow$ mass) and width 
can be given, which connects both descriptions
\begin{eqnarray}
\label{trafo}
R^{(0+1)}(M_W^2)  
& \rightarrow & \overline{R}^{(0+1)}(M_W^2) = 
R^{(0+1)}(M_W^2) \frac{(1-i \gamma)}{(1+\gamma^2)}
\nonumber\\
M_W & \rightarrow & \overline{M}_W = M_W \; (1+\gamma^2)^{-\frac{1}{2}}
\nonumber\\
\Gamma_W^{(0+1)} & \rightarrow & \overline{\Gamma}_W^{(0+1)} = 
\Gamma_W^{(0+1)} \; (1+\gamma^2)^{-\frac{1}{2}} \; .
\end{eqnarray}
Consequently, the $W$ boson mass in the description with $s$-dependent width is
about $\sim 27$ MeV smaller as compared to the constant width approximation.
With the help of these transformations the effect 
of an $s$-dependent width can be easily studied without the necessity to
deal with the - with regard to the $s$-dependence -
complicated scattering amplitude from Eq.~\ref{matp}, especially when
a convolution integral as it is given by Eq.~\ref{falts} has to be calculated.

In recent publications, either in connection with the 
$W$ pair production at LEP II \cite{weakwork}
or with the radiative $W$ production at the Tevatron \cite{baur},
several approaches to consider an $s$-dependent
width in the $W$ propagator in a gauge invariant way 
have been discussed. The prescription given by 
Baur et al.~\cite{baur} results from taking into account the imaginary
part of the virtual fermionic correction to the $\gamma W W$-vertex.
We checked, that applying the transformation we derived
(Eq.~\ref{trafo}) in order to consider an $s$-dependent width
yields the same modification of the bremsstrahlung amplitude 
as presented in \cite{baur}.

\section{The partial $W$ width in ${\cal O}(\alpha^2)$}
\setcounter{equation}{0}\setcounter{footnote}{0}

The partial $W$ width in ${\cal O}(\alpha^2)$
can be written as follows:
\begin{equation}
\Gamma^{(0+1)}_{W\rightarrow f\!f'} = \Gamma^{(0)}_{W\rightarrow f\!f'} 
\; (1+2 \, {\cal R}\!e \; \delta \hat{\Gamma}_{virt.}+\delta \Gamma_{B\!R})
\; ,
\end{equation}
where $\Gamma^{(0)}_{W\rightarrow f\!f'}$ denotes
the partial width in leading order given by Eq.~\ref{gwbmass}.
$\delta \hat{\Gamma}_{virt.}$ and $\delta \Gamma_{B\!R}$
represent the virtual and real contributions, resp., calculated in
$R_{\xi}$ gauge and in the limit of massless decay products.
The discussion of the electroweak ${\cal O}(\alpha)$ contribution 
to the $W$ width performed in Feynman 't Hooft gauge and under
consideration of massive decay products can also be found in \cite{width}.
In the following we concentrate on the gauge invariant separation 
into a QED-like and weak part.  

The Feynman-diagrams representing real photon emission
described by 
\[\delta \Gamma_{B\!R}=
\delta \Gamma^s_{B\!R}+\delta \Gamma^h_{B\!R}\]
are shown in Fig.~8. 
The soft $\delta \Gamma^s_{B\!R}$ and hard $\delta \Gamma^h_{B\!R}$
bremsstrahlung contribution can both be described
by the same form factors we already have derived for
the final state photon emission in the $W$ production process
evaluated at $s=M_W^2$: $\delta \Gamma^s_{B\!R}=F_{B\!R}^{final}(M_W^2)$
given by Eq.~\ref{fbrfin} and $\delta \Gamma^h_{B\!R}$ is defined by 
Eq.~\ref{gammah}.

\begin{figure}
\label{wbr}
\psfig{figure=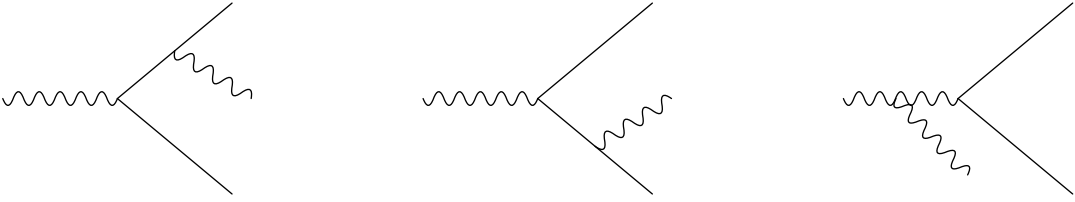}
\caption{Real photon corrections in ${\cal O}(\alpha)$
to the partial $W$ width}
\end{figure}

\begin{figure}
\label{wvirt}
\psfig{figure=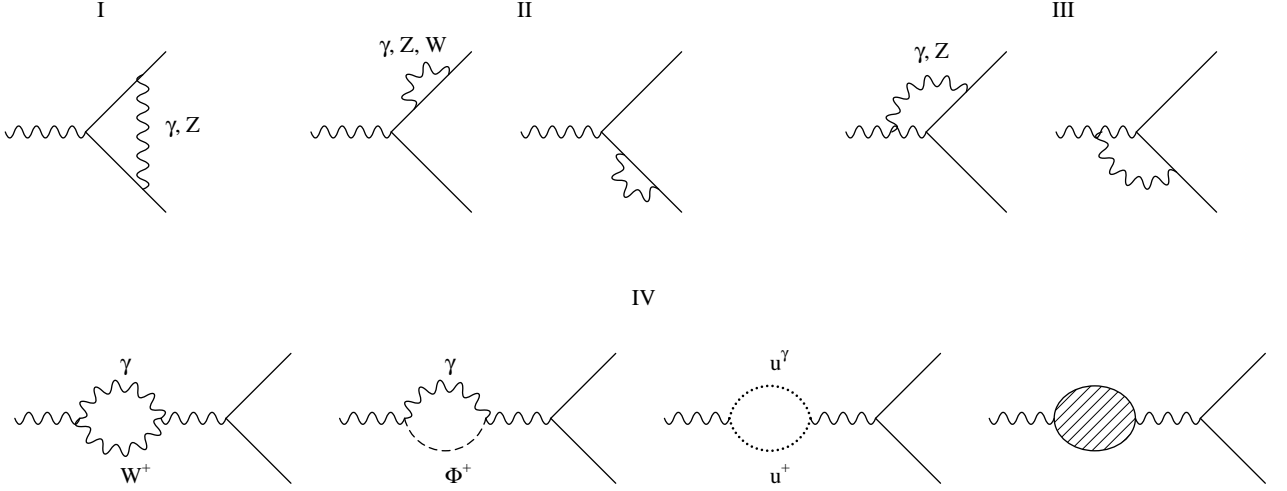}
\caption{Electroweak 1-loop corrections in ${\cal O}(\alpha)$
to the partial $W$ width (again, the non-photonic corrections to the
$W$ self energy are symbolised by the shaded loop)}
\end{figure}

$\delta\hat{\Gamma}_{virt.}$ comprises the renormalised 
vertex correction (diagram I,II,III in Fig.~9 and the counter term 
given by Eq.~\ref{ctv}) and the wave function renormalisation 
for the $W$ boson (diagram IV in Fig.~9 together with Eq.~\ref{cts}). 
Again, we discuss the photon and pure weak contribution separately
\begin{equation}
\label{gamvirt}
\delta\hat{\Gamma}_{virt.}  =  F_{weak}^f(M_W^2)+ F_{\gamma}^f(M_W^2) \; .
\end{equation}
The pure weak contribution can be described by the same
form factor $F_{weak}^f(M_W^2)$ of Eq.~\ref{purewf},
which has been derived from the 
weak corrections to the $W$ decay process of the resonant 
$W$ production in the 4-fermion process.
In contrary, the structure of the virtual 
photon contribution $F_{\gamma}^f(M_W^2)$ differs from that of the
$W$ resonance and requires a separate discussion.
For a $W$ boson being on-shell all photonic 1-loop corrections
in Fig.~9 develop IR-singularities. Thus, in order to gain 
a gauge invariant separation into a QED-like 
$\delta_{Q\!E\!D}^f$ and a (modified) weak part 
$\delta \tilde{\Gamma}_{weak}^f$
\begin{equation}
\label{widthzer}
\Gamma^{(0+1)}_{W\rightarrow f f'} = \Gamma^{(0)}_{W\rightarrow f f'} 
\; (1+2 {\cal R}\!e \; \delta \tilde{\Gamma}_{weak}^f+\delta_{Q\!E\!D}^f)\; ,
\end{equation} 
the diagrams III and IV have also to be considered 
by the YFS-procedure. 
The application of the prescription given in Sec.~3.1 to these diagrams
\begin{equation}
\mbox{diagram III}\; \;  :
\; i\Lambda_{\mu}^{I\!I\!I,f}  =  i g_w \gam (1-\gaf)
\; [F_{I\!I\!I,f}^{I\!R}+F_{I\!I\!I,f}^{finite}](s=M_W^2)
\end{equation}
with
\begin{equation}
\label{dreion}
F_{I\!I\!I,f}^{I\!R}(s=M_W^2) =  (i 4 \pi \alpha)
\left\{\qf \int_D\frac{(2 \pf-k) (k-2q)}{D_\lambda D_f (k^2-2kq)}
+\qfs \int_D\frac{(2 \pfs+k)(k+2q)}{D_\lambda D_{f'} (k^2+2kq)} \right\}
\end{equation}
and
\begin{equation}
\mbox{diagram IV}\; \;  : \; i\Lambda_{\mu}^{I\!V}  =  i g_w \gam (1-\gaf)
\; \frac{1}{2} \; [F_{I\!V}^{I\!R}+F_{I\!V}^{finite}](s=M_W^2)
\end{equation}
with
\begin{equation}
\label{vieron}
F_{I\!V}^{I\!R}(s=M_W^2) =  (i 4 \pi \alpha)
\int_D\frac{(2q-k)^2}{D_\lambda (k^2-2kq)^2} 
\end{equation}
together with the IR-singular parts extracted from the diagrams I,II 
(Eqs.~\ref{ireins},\ref{irzwei} evaluated at $s=M_W^2$)
yield a gauge invariant YFS-form factor multiplying the
tree level $W$ width, which is the same as for the final state 
photon contribution to the $W$ production 
\begin{eqnarray} 
\label{yfswidth}
\lefteqn{\fyfs^{final}(s=M_W^2) = \frac{1}{2} (i 4 \pi \alpha) \int_D
\frac{1}{D_\lambda} \times}
\nonumber\\
& & \left[\overbrace{\frac{\qf (2 \pf-k)_{\rho}}{(k^2-2k\pf)}+
\frac{\qfs (2 \pfs+k)_{\rho}}
{(k^2+2 k\pfs)}-\frac{1}{2} \frac{(\qf-\qfs) (2 q-k)_{\rho}}{(k^2-2 k q)}+
\frac{1}{2}\frac{(\qf-\qfs) (2 q+k)_{\rho}}{(k^2+2 k q)}}^
{k^{\rho}{\cal J}^{final}_{\rho} = 0}\right]^2 
\nonumber\\
& & \mbox{}
\end{eqnarray}
The only difference is, that the ad-hoc
addition of a `zero' in Eq.~\ref{yfsfak} can now be traced back to
the IR-singular contributions of diagrams involving the $\gamma W W$-coupling,
when the $W$ boson is considered to be on-shell.
The explicit expressions for
$F_{I\!I\!I,f}^{I\!R}(M_W^2),F_{I\!V}^{I\!R}(M_W^2)$
and the corresponding IR-finite parts are given by the
Eqs.~\ref{fcwidth}-\ref{fvieriron}. Consequently, 
the QED-like form factor to the $W$ width from Eq.~\ref{widthzer} 
\begin{eqnarray}
\delta^f_{Q\!E\!D} & = & F_{Q\!E\!D}^{final}(s=M_W^2)
+\delta\Gamma_{B\!R}^h
\nonumber\\
& = & \frac{\alpha}{\pi} \; 
\left[\frac{3}{8}\; (\qf^2+\qfs^2)+\frac{7}{3}
+\frac{\pi^2}{24}\right]  
\end{eqnarray}
is the same as for the final state QED contribution to
the $W$ resonance given by Eq.~\ref{sigfin}.
This result can be compared with the `QED-factor' 
for a leptonic $W$ decay given in \cite{brb}
\[\delta_Q = \frac{\alpha}{\pi} \; 
\left[\frac{77}{24}-\frac{\pi^2}{3}\right] \; ,\]
which has been derived by considering from the photonic virtual contribution
only the mass singular logarithms being
gauge invariant by themselves.
 
The IR-finite remnants of the YFS-prescription in the case of the $W$ width
yield
\begin{eqnarray}
\label{wremgam}
\delta \Gamma_{rem.}^{\gamma}&  = & 
\sum_{j=I,I\!I,I\!I\!I} F_{j,f}^{finite}\mid_{subtr.}(M_W^2)
+\frac{1}{2}F_{I\!V}^{finite}(M_W^2)
-\frac{1}{2}\ap (2+\frac{3}{2}\pi^2)
\nonumber\\
& = & \frac{1}{2} \ap \left\{\frac{25}{3}\;  
\dmw+\frac{68}{9}-\frac{3}{2}\pi^2+(\xi_W-1) \alpha_W \right\}-
\frac{1}{2} \delta Z_2^{W,\gamma}  \; ,
\end{eqnarray}
which can be absorbed in a modified weak contribution 
\begin{equation}
\label{modwweak}
\delta\tilde{\Gamma}_{weak}^f = F_{weak}^f(M_W^2)+\delta\Gamma_{rem.}^{\gamma}
\; .
\end{equation}
This completes the gauge invariant separation
of the electroweak corrections in 
${\cal O}(\alpha)$ to the partial $W$ width due to 
Eq.~\ref{widthzer}. Finally, a QED-subtracted partial $W$ width
can be defined 
\begin{equation}
\label{modwidth}
\tilde{\Gamma}^{(0+1)}_{W\rightarrow f\!f'} =
\Gamma^{(0)}_{W\rightarrow f\!f'} 
\; (1+2 \; {\cal R}\!e \, \delta \tilde{\Gamma}_{weak}^f)\; ,
\end{equation}
which will appear in the residue of the Breit-Wigner-form of the 
resonant $W$ production cross section.

\newpage

\section{Feynman-rules}
\setcounter{equation}{0}\setcounter{footnote}{0}

In the following the Feynman-rules, which differ from the ones in
Feynman-'t Hooft-gauge ($\xi_i=1$) are explicitly given.
The remaining Feynman-rules can be found in \cite{spiess}.

\begin{minipage}{7cm}
{\vspace*{1.25cm} \psfig{figure=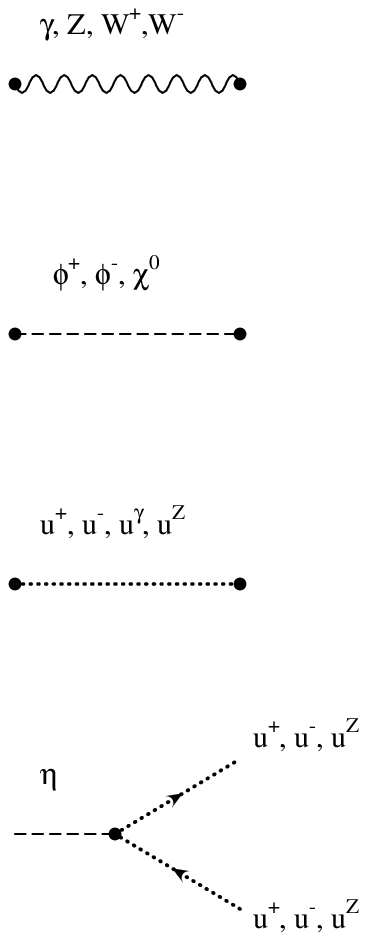}}
\end{minipage}\hfill
\parbox{8cm}{\begin{eqnarray*}
& & \\
& : & \frac{-i}{q^2-M_V^2+i\eps}\; \left(g^{\mu\nu}+\frac{(\xi_V-1)\, q^{\mu}
q^{\nu}}{q^2-\xi_V\; M_V^2}\right)\\
& & \\
& & \\
& & \\
& : & \frac{i}{q^2-\xi_{(W,Z)}\; M_{(W,Z)}^2+i\eps}\\
& & \\
& & \\
& & \\
& : & \frac{i}{q^2-\xi_{(W,Z,\gamma)}\; M_{(W,Z,\gamma)}^2+i\eps}\\
& & \\
& & \\
& & \\
& : & -i\, e \; \frac{M_W}{2 \, s_w}\; [\xi_W;\frac{\xi_Z}{c_w^2}]
\end{eqnarray*}}

As it has already been pointed out, a renormalisation procedure needed
to be performed in order to cope with the arising UV-divergences. Thus,
after the multiplicative renormalisation of the 
$SU(2)$ gauge coupling constant and the gauge boson field $W_{\mu}^a$,
the $Wff'$-vertex counter term yields as follows \cite{spiess}:\\
\begin{minipage}{7cm}
{\psfig{figure=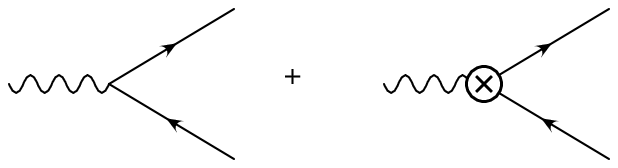}}
\end{minipage}\hfill
\parbox{8cm}{
\begin{equation}
\label{ctv}
: \;  \frac{i e}{2 \sqrt{2} s_w}\gam (1-\gaf) 
(1+\delta Z_1^W-\delta Z_2^W)\end{equation}}\\
and the renormalised $W$ self energy is defined by
\begin{equation}
\label{cts}
\hat\Sigma_T^W(s) = \Sigma_T^W(s) + (s-M_W^2) \; \delta Z_2^W-\delta M_W^2 \; .
\end{equation}
The renormalisation constants determined in 
the on-shell renormalisation scheme are given by \cite{spiess}, \cite{holl}
\begin{eqnarray}
\label{renw}
\delta Z_1^W & = & -\Pi^{\gamma}(0)-\frac{3-2 s_w^2}{s_w c_w}
\; \frac{\Sigma_T^{\gamma Z}(0)}{M_Z^2}+
\frac{c_w^2}{s_w^2} \left[\frac{\delta M_Z^2}{M_Z^2}
-\frac{\delta M_W^2}{M_W^2}\right]
\nonumber\\
\delta Z_2^W & = & -\Pi^{\gamma}(0)-2 \frac{c_w}{s_w}
\; \frac{\Sigma_T^{\gamma Z}(0)}{M_Z^2}+
\frac{c_w^2}{s_w^2} \left[\frac{\delta M_Z^2}{M_Z^2}
-\frac{\delta M_W^2}{M_W^2}\right] 
\end{eqnarray}
with
\begin{equation}
\label{renmass}
\delta M_{(W,Z)}^2 = {\cal R}\!e \Sigma_T^{(W,Z)}(s=M_{(W,Z)}^2) \; .
\end{equation}
$\Pi^{\gamma}$, $\Sigma^{\gamma Z}_T$ denote the photon vacuum polarisation
and the photon-$Z$-mixing, respectively. 

It should be mentioned, that we do not perform an 
`explicit' wave function renormalisation for the
external fermions, but rather take into account the modification due to their
self interaction by the consideration of the 1-loop contributions
shown in Fig.~2 (diagram II). Therefore no renormalisation constant
for the fermion doublet $\delta Z_L$ occurs in the counter term for
the $Wff'$-vertex.

\section{The form factors}   
\setcounter{equation}{0}\setcounter{footnote}{0}

In the following we provide the explicit expressions
for the different contributions
to the form factor describing the virtual electroweak ${\cal O}(\alpha)$
contribution to the $W$ production process 
$\hat F_{virt.}(s,t)$ given by Eq.~\ref{virtsep}.
They are calculated in $R_{\xi}$-gauge, where, following \cite{sir}, the 
$\xi_i$-dependent parts are expressed in terms of the functions
$\alpha_i,v_{ij},\eta_{ij}$. The latter are described in App.~F, where
the explicit expressions for the IR- and/or on-shell singular
scalar 2-,3- and 4-point integrals $B_0,C_0,D_0$ can be found, too.
In order to regularise the arising
IR-singularities a fictive photon mass $\lambda$
has been used. After dimensional regularisation the UV-divergences 
have been extracted in form of the following singular terms: 
\[\Delta_{s} \equiv \Delta-\lo{s}{\mu^2}\right) \; \mbox{ and }\;
\Delta_{m} \equiv \Delta-\lo{m^2}{\mu^2}\right)\]
with $\Delta=\frac{\textstyle 2}{\textstyle{4-D}}-\gamma_E
+\log4\pi$ ($\gamma_E$: Euler constant).

\subsection{The form factor describing the photonic 1-loop corrections}

The photonic form factor $F_{\gamma}(s,t)$ of Eq.~\ref{virtsep} 
is composed as follows:
\begin{equation}
\label{photvirt}
F_{\gamma}(s,t) = 
\sum_{j=I,I\!I,I\!I\!I} F_j^{\gamma}(s)
\underbrace{-\frac{\Sigma_T^{W,\gamma}(s)
-{\cal R}\!e \Sigma_T^{W,\gamma}(M_W^2)}{s-M_W^2}
-\delta Z_2^{W,\gamma}}_{=: F_{I\!V}^{\gamma}(s)}+ F_V^{\gamma}(s,t)
\end{equation}
with
\begin{eqnarray}
F_j^{\gamma}(s) &= & (F_{j,f}^{\gamma}+F_{j,i}^{\gamma})(s)
\nonumber \\
F_V^{\gamma}(s,t) & = & (F_V^t+F_V^u)(s,t) \; .
\end{eqnarray}
In the following the explicit expressions for the 
different contributions to the photonic form factor will be
provided, starting with the
final state photonic vertex corrections. By applying the
substitution $(f,f')\rightarrow (i,i')$ the corresponding intial state
contribution can be easily derived. \\
\underline{{\bf diagram I}}:
\begin{eqnarray}
\label{formfa}
F_{I,f}^{\gamma}(s) 
& = & \ap \qf\qfs\; \left\{
-2s\; C_0(s,\mf,\mfs,\lambda)+2 B_0(\pf^2,\lambda,\mf) 
+2 B_0(\pfs^2,\lambda,\mfs)  \right.
\nonumber\\
&-& \left. 3 B_0(s,\mf,\mfs)-2
+(\xi_\gamma-1) \alpha_\gamma  \right\} \; .
\end{eqnarray}
Performing the loop-integration of Eq.~\ref{ireins} 
the IR-singular contribution is given by
\begin{eqnarray}
F_{I,f}^{I\!R}(s) & = & \ap  \qf\qfs \left\{-2s\; C_0(s,\mf,\mfs,\lambda)
+B_0(\pf^2,\lambda,\mf) +B_0(\pfs^2,\lambda,\mfs)
\right. \nonumber\\
&-& \left.  B_0(s,\mf,\mfs)\right\} 
\nonumber\\
& = & \ap \qf\qfs\left\{\ds+2\lo{s}{\mf\mfs}\right)+2\lo{s}{\mf\mfs}\right)
\lo{\lambda^2}{s}\right)+2
\right. \nonumber\\
&+& \left. \frac{1}{2}\lot{s}{\mf^2}\right)
+\frac{1}{2}\lot{s}{\mfs^2}\right)+\frac{4}{3}\pi^2+i\pi 
[2\lo{s}{\lambda^2}\right)-1]\right\}  \; .
\end{eqnarray}
\underline{{\bf diagram II}}:
\begin{eqnarray}
\label{formfb}
F_{I\!I,f}^{\gamma}(s)
& = & -\frac{1}{2} \ap \;  \left\{\qf^2\; 
\left[\ds+3 \lo{s}{m_f^2}\right)+4+2\; \lo{\lambda^2}{s}\right)+
(\xi_\gamma-1) \alpha_\gamma \right] \right.
\nonumber\\
&+& \left.\qfs^2\; [f\rightarrow f'] \right\} \; .
\end{eqnarray}
Computing the 1-loop integral of Eq.~\ref{irzwei} leads to
\begin{equation}
F_{I\!I,f}^{I\!R}(s) = -\frac{1}{2} \ap \;  \left\{\qf^2\; 
\left[\ds+3 \lo{s}{m_f^2}\right)+4+2\; \lo{\lambda^2}{s}\right)\right]
+\qfs^2\; [f\rightarrow f'] \right\}\; .
\end{equation}
\underline{{\bf diagram V}}:
\begin{eqnarray}
F_{V}^t(s,t)
& = & \ap \; \left\{\qi \qf \left[
-2t(s-M_W^2)\; D_0(s,t,\mi,\mf,M_W,\lambda)+
\frac{(s-M_W^2)}{(s+t)^2}\; f_{V,t}(s,t)\right]\right.
\nonumber\\
&+&\left. \qis \qfs \; [(i,f)\rightarrow(i',f')]\right\} \; .
\end{eqnarray}
In order to provide a complete representation of the 1-loop
corrections also the 
non-resonant contribution $f_{V,t}(s,t)$, which is negligible
in the vicinity of the resonance, will be explicitly given
\begin{eqnarray}
f_{V,t}(s,t) &=& 2(s+t)\; [B_0(s,\lambda,M_W)-B_0(t,\mi,\mf)]
-t (2t+s+M_W^2)\; C_0(1)
\nonumber\\
&+&((s+t)^2+t^2-s M_W^2)\; [C_0(3)+C_0(4)]
+t (s+M_W^2+2t)\; \left[(s-M_W^2)D_0
\right. \nonumber\\
&-& \left .C_0(2)\right]
+(s+t)^2 \; [(\xi_W-1)\eta_{W\gamma}(s)+(\gamma \leftrightarrow W)]\; .
\end{eqnarray}
From Eq.~\ref{boxir} the $t$-channel box contribution to the  
IR singular YFS-form factor is given by
\begin{eqnarray}
F_{V,t}^{I\!R}(s,t)&  = & \ap \; \left\{\qi\qf\; \left[
-2t\; C_0(2)-B_0(t,\mi,\mf)+B_0(\pf^2,\lambda,\mf) 
+B_0(\pii^2,\lambda,\mi)\right] \right.
\nonumber\\
&+& \left. \qis\qfs\;  [(i,f)\rightarrow(i',f')]\right\} 
\nonumber\\
& = & \ap\;  \left\{\qi\qf\left[\lo{t^2}
{\mf^2 \mi^2}\right) 
\lo{\lambda^2}{s}\right)+\frac{1}{2}\lot{s}{\mf^2}\right)+
\frac{1}{2}\lot{s}{\mi^2}\right)
\right.\right.\nonumber\\
&-& \left. \left. \frac{1}{4}\lot{t^2}{s^2}\right)+\frac{\pi^2}{3}
+\ds+\frac{1}{2}\lo{t^2}{s^2}\right)+2\lo{s}{\mf \mi}\right)+2+i \pi\right]
\right. \nonumber \\
&+& \left. \qis\qfs  [(f,i)\rightarrow (f',i')] \right\} \; .
\end{eqnarray}
The application of the substitution described by Eq.~\ref{cross}
leads to the corresponding $u$-channel form factors.

From these IR-singular photonic 1-loop contributions
the following gauge invariant YFS-form factors of Eq.~\ref{yfsfak}
have been extracted:
\begin{eqnarray}
\label{yfsfin}
\fyfs^{final}(s) & = & (F_{I,f}^{I\!R}+F_{I\!I,f}^{I\!R})(s)
\nonumber\\
&+ & \left. \qf (\qf-\qfs) \left[\lo{s}{\mf}\right) \lo{\lambda^2}{s}\right)+
\frac{1}{2} \lot{s}{\mf^2}\right)+\lo{s}{\mf^2}\right)\right. \right.
\nonumber \\
&+ & \left. \left.\ds+3-\frac{1}{2} (1-\frac{3}{2} \pi^2)\right]
-\qfs (\qf-\qfs) [f \rightarrow f']
\right. \nonumber \\
&- & \left. (\qf-\qfs)^2\left[\lo{\lambda^2}{s}\right)+\frac{1}{2} \ds+
\frac{1}{2}\right]\right\}
\\ \label{yfsint}
\fyfs^{interf.}(s,t) & = & (F_{V,t}^{I\!R}+F_{V,u}^{I\!R})(s,t) 
\nonumber\\
&-& \left. \qi (\qf-\qfs) \left[\lo{s}{\mi^2}\right)\lo{\lambda^2}{s}\right)
+\frac{1}{2}\lot{s}{\mi^2}\right)+\lo{s}{\mi^2}\right)
\right. \right.\nonumber\\
&+& \left. \left.\ds+3-\frac{1}{2} (1-\frac{3}{2}\pi^2)\right]
+ \qis (\qf-\qfs) \; [i\rightarrow i']
\right. \nonumber\\
&-& \left. \qf(\qi-\qis)\; [i\rightarrow f]+
\qfs (\qi-\qis) \; [i\rightarrow f']
\right. \nonumber\\
&+& \left. (\qi-\qis) (\qf-\qfs)\left[2 \lo{\lambda^2}{s}\right)+\ds+1\right]
\right\} \; .
\end{eqnarray}
The IR-finite remainders of the YFS-prescription are determined by
\begin{eqnarray}
\label{infin}
F_{j,f}^{finite} & = & F_{j,f}^{\gamma}-F_{j,f}^{I\!R}
\nonumber \\
\label{intfin}
F_{V,(t,u)}^{finite} & = & F_V^{(t,u)}-F_{V,(t,u)}^{I\!R} \; .
\end{eqnarray}

The remaining photonic Feynman-diagrams shown in Fig.~2 
are IR-finite and, thus, are not considered by the YFS-prescription, but
develop on-shell singularities in the vicinity of the $W$ resonance.
In detail, they are described by the following form factors:\\
\underline{{\bf diagram III}}:
\begin{eqnarray}
F_{I\!I\!I,f}^{\gamma}(s)& = &
\ap \left\{\qf \left[2\; C_0(s,m_f,\lambda,M_W)+2 B_0(\pf^2,\lambda,\mf)+
(2+\frac{M_W^2}{s})B_0(\pf^2,\mfs,M_W)
\right. \right.
\nonumber\\ 
&-& \left. \left. (1+\frac{M_W^2}{s})B_0(s,\lambda,M_W)
+\frac{1}{2}[(\xi_W-1) (v_{W\gamma}(s)+\alpha_W)
+(\gamma \leftrightarrow W)] \right]
\right. \nonumber\\
&-&\left. \qfs [f \rightarrow f'] \right\}
\nonumber\\ 
& = & \ap\left\{\qf\left[3 \ds+2 \lo{s}{\mf^2}\right)
+3+2\lo{s}{\mf^2}\right) \lo{|\dw|}{M_W^2}\right)-\frac{\pi^2}{3}
\right.\right.\nonumber\\
&+& \left.\left.  f_{I\!I\!I,f}(s)
+\frac{1}{2}[(\xi_W-1) (v_{W\gamma}(s)+\alpha_W)
+(\gamma \leftrightarrow W)] \right]-\qfs [f \rightarrow f'] \right\} \;,
\end{eqnarray}
where $f_{I\!I\!I,f}(s)$ can be neglected in the resonance region
($w=\frac{\textstyle{M_W^2}}{\textstyle s}$):
\begin{eqnarray}
\lefteqn{f_{I\!I\!I,f}(s)  =  \ap \qf \left\{ (1-w)\; \left[
1+ (1+w) \lo{|\dw|}{M_W^2}\right)-2 \lo{s}{\mf^2}\right)
\lo{|\dw|}{M_W^2}\right)+\frac{\pi^2}{3}\right] \right.}
\nonumber\\
&-& \left. 2 w \mbox{Sp}(1-w)-w \log^2(w)+\log(w)-i \pi \theta(s-M_W^2)
[1-w^2+2w\lo{s}{\mf^2}\right)]\right\} \; .
\nonumber\\
& & \mbox{}
\end{eqnarray}
\underline{{\bf diagram IV}}:\\
The renormalised $W$ self energy contribution is described by 
$F_{I\!V}^{\gamma}(s)$ of Eq.~\ref{photvirt},
where $\delta Z_2^{W,\gamma}$ denotes the photon contribution to the
wave function renormalisation of the $W$ boson given by Eq.~\ref{renw}.
The photon contribution to the $W$ self energy reads
\begin{eqnarray}
\label{sigwp}
\Sigma_T^{W,\gamma}(s) & = & (-\ap) \left\{
\frac{7}{3} M_W^2\dmw +\frac{5}{3}M_W^2+\frac{2}{9} s
+4 s B_0(s,\lambda,M_W)
\right. \nonumber\\
&+& \left. \frac{4}{3}(s-M_W^2) B_1(s,\lambda,M_W) \right.
\nonumber\\
& - & \left. (s-M_W^2)\;
[(\xi_W-1) (v_{W\gamma}(s)+ \frac{1}{2}(s-M_W^2)\eta_{W\gamma}(s))
+(\gamma \leftrightarrow W)]\right\} \; .
\end{eqnarray}
In App.~A also the imaginary part of $\Sigma_T^{W,\gamma}(s)$
has been carefully studied
\begin{equation}
{\cal I}\!m \Sigma_T^{W,\gamma}(s) = 
(s-M_W^2)\;  \theta(s-M_W^2) \; I(s)
\end{equation}
with
\begin{eqnarray}
\label{imag}
I(s)& = & \ap \; \left\{-4 \pi \, \left[1
+\frac{1}{6}(1-\frac{M_W^2}{s})^2\right]
\right.\nonumber\\
&+& \left. {\cal I}\!m \; [(\xi_W-1) (v_{W\gamma}(s)
+ \frac{1}{2}(s-M_W^2)\eta_{W\gamma}(s))
+(\gamma \leftrightarrow W)]\right\} \; ,
\end{eqnarray}
where $v_{W\gamma}(s)$ and $\eta_{W\gamma}(s)$ 
(Eq.~\ref{eichfkt}) develop imaginary parts, when \cite{sir}
\begin{equation}
\label{imagself}
s \geq  M_W^2 \; , \; (\sqrt{\xi_{\gamma,W}}+1)^2\; M_W^2\; ,\; 
 4 \, \xi_{\gamma,W} \, M_W^2 \; .
\end{equation} 
Using Eq.~\ref{sigwp} the form factor $F_{I\!V}^{\gamma}(s)$ defined
by Eq.~\ref{photvirt} yields
\begin{eqnarray}
F_{I\!V}^{\gamma}(s) &=& \ap \left\{\frac{10}{3} \; \dmw
+\frac{68}{9}-4\lo{|\dw|}{M_W^2}\right)
-[(\xi_W-1) v_{W\gamma}(s)+(\gamma \leftrightarrow W)]
\right. \nonumber\\
&+& \left. f_{I\!V}(s)\right\}-\delta Z_2^{W,\gamma} \; ,
\end{eqnarray}
where 
\begin{eqnarray}
f_{I\!V}(s)& = & (1-w)\left[
\frac{2}{3} (1-w)\; \lo{|\dw|}{M_W^2}\right)-\frac{2}{3}
-[(\xi_W-1)\frac{1}{2 s}\eta_{W\gamma}(s)
+(\gamma \leftrightarrow W)]\right]
\nonumber\\
&+& i\pi \theta(s-M_W^2)\; \left[\frac{2}{3}\frac{\dw^2}{s^2}-4\right]
\end{eqnarray}
again describes a contribution, which vanishes for $s=M_W^2$.
Due to Eq.~\ref{renw} the photon contribution to the renormalisation constant
$\delta Z_2^W=\delta Z_2^{W,\gamma}+\delta Z_2^{W,weak}$ 
is determined by $\delta M_W^2$
\begin{equation}
\delta Z_2^{W,\gamma} = \ap \left(\frac{c_w}{s_w}\right)^2\; 
\left[\frac{19}{3}\; \dmw+\frac{89}{9}\right]\; .
\end{equation}

In the course of the extraction of a gauge invariant YFS-form factor from
the photonic 1-loop corrections to the $W$ width
the IR-singular Feynman-diagrams III and IV of Fig.~5
also needed to be considered.
In the following, we provide the explicit expressions for the
complete form factor $F_{j,f}^{\gamma}(M_W^2)$
and the IR-singular part $F_{j,f}^{I\!R}$ extracted according to the
YFS-prescription, now evaluated at $s=M_W^2$. \\
\underline{{\bf diagram III}}:
\begin{eqnarray}
\label{fcwidth}
F_{I\!I\!I,f}^{\gamma}(M_W^2)
& = & \ap\left\{\qf\left[3 \dmw+4 \lo{M_W}{\mf}\right)
+2 \lot{M_W}{\mf}\right)+3
\right. \right. \nonumber\\
&+& \left. \left. 4\lo{M_W}{\mf}\right) \lo{\lambda}{M_W}\right)
\right. \right. \nonumber\\
&+& \left.\left.  
\frac{1}{2}[(\xi_W-1) (v_{W\gamma}(M_W^2)+\alpha_W)
+(\gamma \leftrightarrow W)] \right]-\qfs [f \rightarrow f'] \right\}
\end{eqnarray} 
Performing the loop-integration in Eq.~\ref{dreion} leads to
\begin{eqnarray}
F_{I\!I\!I,f}^{I\!R}(M_W^2)
& = & \ap\left\{\qf\left[\dmw+2 \lo{M_W}{\mf}\right)
+2 \lot{M_W}{\mf}\right)+3
\right. \right. \nonumber\\
&+& \left. \left. 4\lo{M_W}{\mf}\right) \lo{\lambda}{M_W}\right)\right]
-\qfs [f \rightarrow f'] \right\} \; .
\end{eqnarray} 
\underline{{\bf diagram IV}}:
\begin{eqnarray}
\label{fdwidth}
F_{I\!V}^{\gamma}(M_W^2) & = &
- \lim_{s\rightarrow M_W^2}
\frac{\Sigma_T^{W,\gamma}(s)-{\cal R}\!e \Sigma_T^{W,\gamma}(M_W^2)}
{s-M_W^2}-\delta Z_2^{W,\gamma}
\nonumber\\
& = & -\left. \frac{\partial \Sigma_T^{W,\gamma}(s)}{\partial s}
\right|_{s=M_W^2}
-\delta Z_2^{W,\gamma}
\end{eqnarray}
Using Eq.~\ref{sigwp} $F_{I\!V}^{\gamma}(M_W^2)$ is given by
\begin{equation}
\label{wselfon}
F_{I\!V}^{\gamma}(M_W^2) = \ap \left\{\frac{10}{3} \; \dmw
+\frac{32}{9}-4\lo{\lambda}{M_W}\right)
-[(\xi_W-1) v_{W\gamma}(M_W^2)+(\gamma \leftrightarrow W)]\right\}
-\delta Z_2^{W,\gamma} \; .
\end{equation}
The explicit expression for Eq.~\ref{vieron} reads
\begin{equation}
\label{fvieriron}
F_{I\!V}^{I\!R}(M_W^2) = -\ap \left\{\dmw+4+4\lo{\lambda}{M_W}\right)\right\}
\; .
\end{equation}

\subsection{The form factors describing the pure weak 1-loop corrections}

The pure weak form factor $F_{weak}(s=M_W^2)$
is given by Eq.~\ref{wres} with the final state contribution
\begin{equation}
\label{purewf}
F_{weak}^f(M_W^2) = 
\sum_{j=I,I\!I,I\!I\!I} F_{j,f}^{weak}(M_W^2)+\delta Z_1^W-\delta Z_2^W
\underbrace{-\frac{1}{2}\frac{\partial\Sigma_T^{W,weak}(s)}
{\partial s}|_{s=M_W^2}
-\frac{1}{2}\delta Z_2^{W,weak}}_{=: F_{I\!V,f}^{weak}(M_W^2)} \: .
\end{equation}
Performing the substitution $(f,f')\rightarrow (i,i')$ yields
the corresponding initial state form factor $F_{weak}^i(M_W^2)$.
In the following we provide the explicit 
expressions for the different contributions in Eq.~\ref{purewf}.\\
\underline{{\bf diagram I}}($Z$-boson exchange):
\begin{eqnarray}
\lefteqn{F_{I,f}^{weak}(s)  =  \ap (v_f+a_f)(v_{f'}+a_{f'})
\left\{\dmz-(2z+3)\log(z)-2z-4 \right.}
\nonumber\\
&+& \left. 2(1+z)^2\; \left[\log(z)\lo{z+1}{z}\right)
-\mbox{Sp}(-\frac{1}{z})\right]-i\pi [2z+3+2(1+z)^2
\lo{1+z}{z}\right)]
\right. \nonumber\\
&+& \left. (\xi_Z-1)\alpha_Z \right\}
\end{eqnarray}
with $z=\frac{\textstyle{M_Z^2}}{\textstyle s}$ and the couplings
$v_f=(I_3^f-2s_w^2\qf)/(2 s_w c_w)$, $a_f=I_3^f/(2s_w c_w)$.\\
\underline{{\bf diagram II}}($Z$- and $W$-boson exchange):
\begin{equation}
F_{I\!I,f}^{weak}(s)=F_{I\!I,f}^Z(s)+F_{I\!I,f}^W(s)
\end{equation}
with
\begin{equation}
F_{I\!I,f}^Z(s)  =  \frac{1}{2}\ap [(v_f+a_f)^2+(v_{f'}+a_{f'})^2]\; 
\left\{-\dmz+\frac{1}{2}-(\xi_Z-1)\alpha_Z\right\}
\end{equation}
\begin{equation}
F_{I\!I,f}^W(s)  =  \frac{1}{2}\ap \frac{1}{s_w^2}
\left\{-\dmw+\frac{1}{2}-(\xi_W-1)\alpha_W\right\} \; .
\end{equation}
\underline{{\bf diagram III}}($Z$-boson exchange):
\begin{eqnarray}
F_{I\!I\!I,f}^{weak}(s) & = & \ap \frac{c_w}{s_w} (v_f+a_f-v_{f'}-a_{f'})
\left\{\frac{1}{2} (4+w+z) (\dmz+\dmw)+(w-z)\lo{M_Z}{M_W}\right)
\right.\nonumber\\
&-& \left. (w+z+1) B_0(s,M_Z,M_W)+
2 s (z+w+wz) \, C_0(s,m_{(f,f')}=0,M_W,M_Z)
\right. \nonumber\\
&+& \left. 4+w+z+\frac{1}{2} [(\xi_W-1)[v_{WZ}(s)+\alpha_W]+
(W\rightarrow Z)]\right\} \; .
\end{eqnarray}
The scalar 3-point integral $C_0$ evaluated at $s=M_W^2$ 
yields as follows:
\begin{equation}
C_0(s=M_W^2,0,M_W,M_Z) = -\frac{1}{M_W^2}\; \lo{x_1}{x_1-1}\right)
\, \lo{x_2}{x_2-1}\right)
\end{equation}
with
\[ x_{1,2} = \frac{M_Z^2}{2 M_W^2}\; \left(1\pm i \,
 \sqrt{\frac{4 M_W^2}{M_Z^2}-1}\right) \; .\]
\underline{{\bf vertex counter part}}:\\
The explicit expression for the counter part to the  
$Wff'$-vertex (Eq.~\ref{ctv}) reads as follows:
\begin{equation}
\delta Z_1^W-\delta Z_2^W = \ap \frac{1}{s_w^2}\;
(-2\dmw-(\xi_W-1) v_W(0))\; .
\end{equation}
\underline{{\bf diagram IV}}:\\
The contribution of the renormalised $W$ self energy 
to the weak form factor $F_{I\!V,f}^{weak}(M_W^2)$ of Eq.~\ref{purewf}
is determined by
\begin{eqnarray}
\delta Z_2^{W,weak} & = &
\ap \left\{-\frac{4}{3} \sum_f \qf^2\Delta_{\mf}+
(3-4 \left(\frac{c_w}{s_w}\right)^2)\dmw+\frac{2}{3}
-\frac{2}{s_w^2}(\xi_W-1) v_W(0)\right\}
\nonumber\\
&+& \left(\frac{c_w}{s_w}\right)^2 \left[
\frac{{\cal R}\!e \Sigma_T^Z(M_Z^2)}{M_Z^2}
-\frac{{\cal R}\!e \Sigma_T^{W,weak}(M_W^2)}{M_W^2}\right]
\end{eqnarray}
and the derivative of $\Sigma_T^{W,weak}$ given by Eq.~\ref{wselfx},
\ref{sigwfull}.
The $\xi_i$-dependence of the 
$Z$ self energy and the weak 1-loop correction to the 
$W$ self energy reads as follows
($(v,\eta)_{i,j}\stackrel{i=j}{\equiv}(v,\eta)_i$):
\begin{eqnarray}
\label{zselfx}
\Sigma_T^Z(s)& = & \Sigma_T^Z(s)|_{\xi_i=1}
\nonumber\\
&+& \ap 2 \; \frac{c_w^2}{s_w^2} (s-M_Z^2)\; (\xi_W-1)\; 
[v_W(s)+\frac{1}{2}(s-M_Z^2)\eta_W(s)]
\\
\label{wselfx}
\Sigma_T^{W,weak}(s)& = &  \Sigma_T^{W,weak}(s)|_{\xi_i=1}
\nonumber\\
&+& \ap \frac{c_w^2}{s_w^2} (s-M_W^2)\; 
[(\xi_W-1) [v_{WZ}(s)+\frac{1}{2}(s-M_W^2)\eta_{WZ}(s)]
+(W\leftrightarrow Z)] \; ,
\nonumber\\
& & \mbox{}
\end{eqnarray}
so that, finally, the $\xi_i$-dependent part of the weak form factor
yields
\begin{equation}
\label{xiweak}
F_{weak}^f(M_W^2) = F_{weak}^f(M_W^2)|_{\xi_i=1}-\frac{1}{2}\ap 
\;(\xi_W-1)\; \alpha_W \; ,
\end{equation}   
which cancels the 
$\xi_W$-dependence of the IR-finite photonic correction
$\delta\Gamma_{rem.}^{\gamma}$ from Eq.~\ref{formweakm}.

For the sake of completeness the explicit expressions for the
$Z$ self energy and the non-photonic contribution to the $W$ self energy
in Feynman-'t Hooft gauge 
will also be provided, although they are already given in \cite{spiess}:
\begin{eqnarray}
\Sigma_T^{Z}(s)|_{\xi_i=1} &  = & \ap \; \left\{
\sum_{f\neq\nu} \frac{4}{3} N_c^f \left[(v_f^2+a_f^2)
\; (s \Delta_{m_f}+(2 \mf^2+s)\; F(s,\mf,\mf)
-\frac{s}{3}) 
\right. \right. \nonumber\\
&-& \left.\left. a_f^2 6 \mf^2\; (\Delta_{m_f}
+F(s,\mf,\mf))\right]
+\sum_{f=\nu} \frac{8}{3} a_f^2 s \; 
\left[\Delta-\lo{s}{\mu^2}\right)+\frac{5}{3}\right]
\right. \nonumber\\
&+& \left.  \frac{1}{c_w^2 \, s_w^2} \; \left[
(-(12 c_w^4-4 c_w^2+1) \; B_2^0
+2 (-2 s c_w^4-2 M_W^2 c_w^2+M_W^2)\;  B_0)(s,M_W,M_W)
\right. \right. \nonumber\\
& + & \left. \left. (6 c_w^4-2 c_w^2+\frac{1}{2}) \; A_0(M_W)
-\frac{2}{3} \; s \; c_w^4
\right. \right. \nonumber\\
&+ & \left. \left.  (-B_2^0+M_Z^2 \; B_0)(s,M_Z,M_{\eta})
+\frac{1}{4}\; (A_0(M_{\eta})+ A_0(M_Z))\right]\right\}
\end{eqnarray}
and
\begin{eqnarray}
\label{sigwfull}
\Sigma_T^{W,weak}(s)|_{\xi_i=1} & = & \ap \frac{1}{s_w^2}\left\{
\frac{1}{3} \sum_{f=e,\mu,\tau} \left[(s-\frac{3}{2} m_f^2)\;
\Delta_{m_f}
 + (s-\frac{m_f^2}{2}-\frac{m_f^4}{2s})\; F(s,0,\mf)
+\frac{2}{3} s-\frac{m_f^2}{2} \right]   
\right. \nonumber\\
&+& \left. \sum_{(q_{+},q_{-})} N_c^f \left[(2 B_2^0+\frac{1}{2}
(s-m_+^2-m_-^2) \;  B_0)(s,m_+,m_-)
\right. \right. \nonumber\\  
& - & \left. \left. \frac{1}{2} (A_0(m_+)+A_0(m_-))\right]
\right. \nonumber\\
&+& \left. [-(8 c_w^2+1)\; B_2^0+(s_w^4 M_Z^2-c_w^2(4 s+M_W^2+M_Z^2))
\; B_0](s,M_Z,M_W)
\right. \nonumber\\
& + & \left. (2 c_w^2+\frac{1}{4}) A_0(M_Z)+
(-c_w^2+\frac{7}{2})A_0(M_W)-2 M_W^2+2 c_w^2 (M_W^2-\frac{1}{3}s)
\right. \nonumber\\
& + & \left. [-B_2^0+M_W^2 B_0](s,M_W,M_{\eta})
+\frac{1}{4} A_0(M_{\eta})\right\} 
\end{eqnarray}
with
\begin{eqnarray}
B_2^0(s,m_1,m_2) & = & \frac{1}{3}\; \left[
(m_1^2 \, B_0+\frac{1}{2}(s+m_1^2-m_2^2) B_1)(s,m_1,m_2)
\right. \nonumber\\
& + & \left. \frac{1}{2} \; A_0(m_2)+\frac{m_1^2+m_2^2}{2}
-\frac{s}{6}\right]
\\
A_0(m)& = & m^2\; \left(\Delta_m+1 \right) \; .
\end{eqnarray}   
The function $F(s,m_1,m_2)$ can be found in \cite{spiess}.

\subsection{The form factor describing the soft photon radiation}

Performing the photon phase space integration in Eq.~\ref{brinv}
leads to the following gauge invariant form factors in the
soft photon limit:
\begin{eqnarray}
\label{fbrin}
F_{B\!R}^{initial}(s) & = & \ap\left\{\qi\qis\left[8 \lo{s}{\mis \mi}\right)
\; [\lw+\delta_p(s)]-\lot{s}{\mi^2}\right)
-\lot{s}{\mis^2}\right)-\frac{4}{3}\pi^2\right]
\right. \nonumber\\
&-& \left. 2 \; \qi^2\left[2 [\lw+\delta_p(s)]
-\lo{s}{\mi^2}\right)\right]-2 \;\qis^2 \; [i\rightarrow i']
\right. \nonumber\\
&+& \left. 2 \; \qi (\qi-\qis) \left[2 \lo{s}{\mi^2}\right)
\; [\lw+\delta_p(s)]-\frac{1}{2}\lot{s}{\mi^2}\right)
-\frac{\pi^2}{3}\right]
\right. \nonumber\\
&-& \left. 2 \; \qis (\qi-\qis) \; [i\rightarrow i']-
4\; (\qi-\qis)^2 \left[\lw+\delta_p(s)-1 \right]
\right\} 
\end{eqnarray}
with 
\[\lw\equiv\lo{2 \Delta E}{\lambda}\left|\frac{\dw}
{\dw-2 \sqrt{s} \Delta E}\right|\right) \]
and $\delta_p$ from Eq.~\ref{phaseshift},
\begin{equation}
\label{fbrfin}
F_{B\!R}^{final}(s) = F_{B\!R}^{initial}(s) \; \; \mbox{ with }\; \; 
[(i,i');\lw,\delta_p] \rightarrow [(f,f'); \lo{2 \Delta E}{\lambda}\right);0]
\end{equation}
and
\begin{eqnarray}
\label{fbrint}
F_{B\!R}^{interf.}(s,t) & = & \ap\left\{\qi\qf\left[4 \lo{t^2}
{\mf^2 \mi^2}\right) \; \lw-\lot{s}{\mf^2}\right)-
\lot{s}{\mi^2}\right) \right. \right.
\nonumber\\
&-& \left. \left. 4\mbox{Sp}\left(1+\frac{s}{t}\right)-\frac{4}{3}
\pi^2\right]
\right. \nonumber \\
&+& \left. 2\; \qis\qfs \; [(f,i)\rightarrow (f',i')]
-2 \; \qis\qf\; [(i,t)\rightarrow(i',u)]
\right. \nonumber\\
&-&\left. 2 \; \qi\qfs\; [(f,t)\rightarrow(f',u)]
\right. \nonumber\\
&-& \left. 2 \; \qi (\qf-\qfs) \left[2 \lo{s}{\mi^2}\right) \;  \lw
-\frac{1}{2}\lot{s}{\mi^2}\right)-\frac{\pi^2}{3}\right]
\right. \nonumber\\
&+&\left. 2 \; \qis (\qf-\qfs)\;  [i\rightarrow i']
-2 \;\qf(\qi-\qis) \; [i\rightarrow f]
\right. \nonumber\\
&+& \left. 2 \; \qfs (\qi-\qis) \; [i\rightarrow f']
+ 8 \; (\qi-\qis) (\qf-\qfs)\left[\lw-1 \right]
\right\}\; .
\end{eqnarray}

\section{The hard photon contribution}
\setcounter{equation}{0}\setcounter{footnote}{0}

The differential cross section for the process 
$i(\pii) i'(\pis) \rightarrow f(\pf) f'(\pfs) \gamma(k)$ reads
in the CMS as follows (with $s=(q^0)^2$ and $q^0$ denotes the
CM energy):
\begin{equation}
d\sigma_h = \frac{1}{2s}\frac{1}{(2\pi)^5}\; 
\frac{d^3 \pf\; d^3 \pfs \; d^3 k}{8 \pf^0\pfs^0 k ^0}
\; \delta(\pii+\pis-\pf-\pfs-k) \; \overline{\sum} |{\cal M}_{B\!R}|^2 \; ,
\end{equation}
where the matrix element ${\cal M}_{B\!R}$
results from the application of the MSM Feynman-rules to the
bremsstrahlung diagrams shown in Fig.~3 
(now without any restriction on the photon momentum $k$;
$\dw=s-M_W^2$) 
\begin{eqnarray}
{\cal M}_{B\!R}&  = & i \frac{\pi \alpha}{2 s_w^2}
 \; \sqrt{4\pi\alpha}\;  \frac{1}{\dw}
\;  \left\{\overline{u}_f\,  G_{\mu,f}^{\rho}\, (1-\gaf) \,
v_{f'}\, \overline{v}_{i'}
\gamma^{\mu}(1-\gaf)u_i
\right. \nonumber\\
& -& \left. \frac{\dw}{\dw-2kq}\; [\overline{u}_f \, 
\gam(1-\gaf)\, v_{f'} \, \overline{v}_{i'} \,
G_i^{\mu\rho}(1-\gaf)\, u_i]\right\} \; \eps^{\ast}_{\rho}(k) \; ,
\end{eqnarray}
where $\eps_{\rho}$ denotes the photon polarisation vector and
\begin{eqnarray}
\label{gmrf}
G^{\mu\rho}_f & = & 
\qf\, \frac{(\pf^{\rho}+\gamma^{\rho}\,  \ks/2)\, \gamma^{\mu}}
{k\pf}-\qfs\, \frac{\gamma^{\mu}\, (\pfs^{\rho}+\ks\, \gamma^{\rho}/2)}
{k\pfs}-\frac{\gamma^{\mu}\, q^{\rho}+k^{\mu}\, \gamma^{\rho}
-g^{\mu\rho}\, \ks}{kq}
\nonumber\\
G^{\mu\rho}_i & = & \qi\, \frac{\gamma^{\mu}\, (\pii^{\rho}
-\ks\, \gamma^{\rho}/2)}
{k\pii}-\qis\, \frac{(\pis^{\rho}-\gamma^{\rho}\, \ks/2)\, \gamma^{\mu}}
{k\pis}-\frac{\gamma^{\mu}\, q^{\rho}-k^{\mu}\,  \gamma^{\rho}
+g^{\mu\rho}\, \ks}{kq} \: .
\nonumber\\
& & \mbox{}
\end{eqnarray}
The initial and final state currents are separately conserved:
$k_{\rho} \, G^{\mu\rho}_f=(\qf-\qfs-1)\gamma^{\mu}=0$ and
$k_{\rho}\, G^{\mu\rho}_i=(\qi-\qis-1)\gamma^{\mu}=0$. 
At first, the Lorentz-invariant 3-particle phase space
\begin{equation}
I = \int\;  \frac{d^3 \pf\; d^3 \pfs \; d^3 k}{8 \pf^0\; \pfs^0 \; k^0}
\; \delta(\pii+\pis-\pf-\pfs-k) 
\end{equation}
will be thoroughly discussed. Under consideration of the energy momentum 
conservation described by the $\delta$-function
the phase space integration will be rewritten, so that 
only the photon phase space integration survives in order to gain
the photon spectra describing hard photon radiation.  
We follow the procedure suggested in
\cite{hard},\cite{wit} and choose the following coordinate system:
the momenta $\vec{p}_i$ and $\vec{k}$ are in the 
(1,3)-plane, with the photon momentum along the third axis.\\
The spatial part of the $\delta$-function
constraints the momenta in such a way,
that in the CMS ($\vec{q}=\vec{p}_i+\vec{p}_{i'}=0$) the relation
$|\vec{p}_f|=|\vec{p}_{f'}+\vec{k}|=\pf^0$ holds and
the phase space integral can be written as follows 
\begin{equation}
I = 2\pi\;  \int_{\Delta E}^{\omega} \frac{|\vec{k}|\, k^0 \,dk^0}{2\, k^0}
\int_{-1}^{1} dx \int_{p_a}^{p_b} \frac{|\vec{p}_{f'}|\,
\pfs^0\,  d\pfs^0}{2\,  \pfs^0}
\int_0^{2\pi} d \Phi \int_{-1}^1 \frac{dz}{2\, \pf^0}
\delta(\pii^0+\pis^0-\pfs^0-k^0-\pf^0)
\end{equation}
with $x=\cos \angle(\vec{k},\vec{\pii}),z=\cos\angle (\vec{k},\vec{\pfs})$
and $\Phi$ denotes the azimuthal angle of $\pfs$ with respect to
the (1,3)-plane.
Since the soft photon contribution has already been discussed separately
the lower bound of photon phase space integration can be
chosen to be $|\vec{k}|=\Delta E$ and no IR singularities occur.
Using 
\begin{equation}
\delta(f(x)) = \frac{\delta(x-x_0)}{|f'(x)|_{x=x_0}}\; ,
\end{equation}
where $f(x)$ is an arbitrary function with 
$f(x_0)=0$ (here: $f(z)=\pf^0$)
\[\delta(\pii^0+\pis^0-\pfs^0-k^0-\pf^0)
=\left|\frac{\pf^0}{|\vec{k}|\,  |\vec{p}_{f'}|}\right|_{z=z_0} 
\, \delta(z-z_0) \] 
with
\[ 2 |\vec{k}|\, |\vec{p}_{f'}|\, z_0=(q^0-k^0-\pfs^0)^2-(k^0)^2-(\pfs^0)^2
+\mfs^2-\mf^2  \; ,\] 
the phase space integral $I(s)$ can be written as follows:
\begin{equation}
I=\pi \int_{\Delta E}^{\omega} \frac{dk^0}{2} \int_{-1}^1 d x \int_{p_a}^{p_b} 
\frac{d\pfs^0}{2} \int_0^{2\pi} d\Phi  \; .
\end{equation}
The requirement $-1 \leq z_0\leq 1$ leads to the following 
limits on the $p_{f'}^0$-integration:
\begin{eqnarray}
p_{a,b} & =& \frac{(q^0-k^0)\, \kappa \, \pm \, k^0 \, 
\sqrt{(\kappa-2 \mfs^2)^2
-4\, \mfs^2 \, \mf^2}}{2 \, (\kappa-\mfs^2+\mf^2)}
\\
\label{omega}
\omega & = & \frac{(q^0)^2-(\mf+\mfs)^2}{2 \, q^0}
\end{eqnarray}
with
\[\kappa=q^0 \, (q^0-2 k^0)+\mfs^2-\mf^2 \; .\]
Finally, after introducing a new variable $y$
\[\pfs^0=\frac{\kappa}{2\, (q^0-k^0)}+\frac{k^0\,  \pii^0}{q^0}\;  y \; ,\]
the starting point for obtaining the hard photon spectra is reached
(with $\pii^0=q^0/2$)
\begin{equation}
\sigma_h(s) = \frac{1}{16 s}\frac{1}{(2\pi)^4}\; 
\int_{\Delta E}^{\omega} \frac{d k^0 \; k^0}{2}\int_{-1}^1dx\int_{y_a}^{y_b} dy
\int_0^{2\pi} d\Phi 
\; \overline{\sum} |{\cal M}_{B\!R}|^2 \; .
\end{equation}
The computation of the spin averaged squared matrix element leads to 
the initial state, final state and interference contributions depending
only on the scalar products of the involved four momenta, which have to
be expressed in terms of the integration variables, e.g.
\begin{equation}
\pis\,  \pfs = \pis^0 \, \pfs^0+|\vec{p}_{i}|\, |\vec{p}_{f'}| \, \cos\varphi
\end{equation} 
with
\[ \cos\varphi=(x z +\sqrt{1-x^2} \; \sqrt{1-z^2} \; \cos\Phi)|_{z=z_0}\; .\]
Finally, the performance of all integrations up to the one over the
photon energy yields the following hard photon spectra
(with $k=2 k^0/q^0$ and $k_m=2 \Delta E/q^0$):
\begin{eqnarray}
\label{spekin}
\sigma_h^{initial}(s) & = & \tilde{\sigma}^{(0)}(s)
\; \int_{k_m}^1 dk \left|\frac{\dw}{\dw-sk}\right|^2 \frac{1-k}{2k}\; 
\left\{\beta_i(s)\; [1+(1-k)^2]+\frac{\alpha}{\pi}\frac{k^2}{3}\right\}
\end{eqnarray}
\begin{eqnarray}
\label{spekfin}
\sigma_h^{final}(s) & = & \tilde{\sigma}^{(0)}(s)
\; \int_{k_m}^1 \frac{dk}{2k}\; 
\left\{\beta_f(s)\; [1+(1-k)^2]+\frac{\alpha}{\pi}\; \frac{k^2}{3}
\right.\nonumber\\
&+& \left. \frac{\alpha}{\pi}(\qf^2+\qfs^2)\; [1+(1-k)^2]\; \log(1-k)]\right\}
\end{eqnarray}
\begin{eqnarray}
\sigma_h^{interf.}(s) & = & \tilde{\sigma}^{(0)}(s) \frac{\alpha}{\pi}
\; \int_{k_m}^1 \frac{dk}{k}\;\left[\frac{\dw}{\dw-sk}+\frac{\dw^{\ast}}
{\dw^{\ast}-sk}\right]\;  \frac{5}{12} \; \{3k-k^2-2\}
\end{eqnarray}
The final state hard photon spectrum $\sigma_h^{final}(s)$
coincides with the result obtained in \cite{brb}.
From the photon spectra the total cross sections
describing hard photon radiation can be obtained
\begin{eqnarray}
\label{hin}
\sigma_h^{initial}(s) & = & \tilde{\sigma}^{(0)}(s)\;  \beta_i(s)
\left\{\lo{|\dw-2 \sqrt{s}\Delta E|}{2 \sqrt{s} \Delta E}\right)
\right. \nonumber\\
& + & \left.\frac{s-M_W^2}{M_W\Gamma_W^{(0+1)}}
\left[\arctan\left(\frac{M_W}{\Gamma_W^{(0+1)}}\right)-
\arctan\left(\frac{2 \sqrt{s} \Delta E-s+M_W^2}{M_W \Gamma_W^{(0+1)}}\right)
\right]\right\}
\end{eqnarray}
\begin{eqnarray}
\label{hfin}
\sigma_h^{final}(s) & = & \tilde{\sigma}^{(0)}(s) 
\; \left\{\beta_f(s)\; \lo{\sqrt{s}}{2 \Delta E}\right)
\right. \nonumber\\
&+& \left. \frac{\alpha}{\pi}\; \left[\qf^2
\; \left(-\frac{3}{4}\lo{s}{\mf^2}\right)-\frac{\pi^2}{6}
+\frac{11}{8}\right)+\qfs^2\; (f\rightarrow f')
+\frac{5}{6}\right]\right\}
\end{eqnarray}
\begin{eqnarray}
\label{hint}
\sigma_h^{interf.}(s) & = & \tilde{\sigma}^{(0)}(s) \; \frac{\alpha}{\pi}
\frac{1}{3}\; [5(\qi\qf+\qis\qfs)+4(\qis\qf+\qfs\qi)] \;
\lo{2\Delta E \sqrt{s}}{\mid \dw-2 \sqrt{s}\Delta E \mid}\right) \; .
\nonumber\\
& & \mbox{}
\end{eqnarray}
Since we are interested on the contribution in the vicinity 
of the $W$ resonance terms $\propto (s-M_W^2)$ and
$\propto \Delta E$ have been neglected.

The parametrisation of the 3-particle phase space in the course of the  
computation of hard bremsstrahlung for the case of the 
$W$ width is less complicated, since the orientation 
of the dreibein made of the three outgoing momenta 
can be freely chosen: the solid angle $\Omega$ determines the orientation 
of the photon momentum and $\Phi$ describes the rotation of the  
$(\vec{p}_f,\vec{p}_{f'})$-system around $\vec{k}$. 
Thus, the hard photon contribution to the
partial $W$ width (in the CMS of the $W$ boson with $q^2=M_W^2$)
\begin{equation}
d\Gamma^h_{W\rightarrow f\!f'} = \frac{1}{2M_W}\frac{1}{(2\pi)^5}\; 
\frac{d^3 \pf\; d^3 \pfs \; d^3 k}{8 \pf^0\pfs^0 k ^0}
\; \delta(q-\pf-\pfs-k) \; \overline{\sum} |{\cal M}_{B\!R}^{final}|^2 \; ,
\end{equation}
turns into \cite{wit}
\begin{equation}
\Gamma^h_{W\rightarrow f\!f'} =  \frac{1}{2M_W}\frac{1}{256 \pi^5}\; 
\int_{\Delta E}^{\omega} d k^0 \int_0^{4\pi} d\Omega \int_0^{2\pi}
d\Phi \int_{x_-}^{x^+} dx  \overline{\sum} |{\cal M}_{B\!R}^{final}|^2 \; ,
\end{equation}
where $\omega$ is given by Eq.~\ref{omega} and the substitution
$p_{f,f'}=\pm x+(M_W-k^0)/2$) has been performed. 
The limits on the $x$-integration $x_{\pm}$ are given by
\begin{equation}
x_{\pm} = \frac{1}{2 \tilde M} \left\{
\frac{\mf^2-\mfs^2}{2 M_W} (M_W-k^0)\pm k^0 
\sqrt{(\tilde M-\frac{(\mf+\mfs)^2}{2M_W})
(\tilde M-\frac{(\mf-\mfs)^2}{2M_W})} \right\}
\end{equation}
with $\tilde M=M_W/2-k^0$.
The matrix element ${\cal M}_{B\!R}^{final}$ reads as follows
($\eta^{\mu}$: polarisation vector of the $W$ boson):
\begin{equation}
{\cal M}_{B\!R}^{final} = i \, \frac{\sqrt(2) \pi \alpha}{s_w}
\, \overline{u}_f\,  G_{\mu,f}^{\rho}\, (1-\gaf) \,
v_{f'}\, \eta^{\mu}(q) \, \eps^{\ast}_{\rho}(k)
\end{equation}
with $G_{\mu,f}^{\rho}$ given by Eq.~\ref{gmrf}, which leads to the same 
hard photon spectrum as for the case of finale state 
bremsstrahlung in the 4-fermion process
($\rightarrow$ Eq.~\ref{spekfin})
\begin{eqnarray}
\label{gammah}
\Gamma^h_{W\rightarrow f\!f'}& = &\Gamma_{W\rightarrow f\!f'}^{(0)} \; 
\int_{k_m}^1 \frac{dk}{2k}\; \left\{\beta_f(M_W^2)\; [1+(1-k)^2]
\right. \nonumber\\
&+& \left. \frac{\alpha}{\pi}\; \frac{k^2}{3}+
\frac{\alpha}{\pi}(\qf^2+\qfs^2)\; [1+(1-k)^2]\; \log(1-k)]\right\}
\nonumber\\
&=:& \Gamma_{W\rightarrow f\!f'}^{(0)} \; \delta\Gamma_{B\!R}^h \; .
\end{eqnarray}
Thus, the factor $\delta\Gamma_{B\!R}^h$ coincides with the one, which 
multiplies the Born-cross section in Eq.~\ref{hfin} evaluated at $s=M_W^2$.

\section{Integrals}
\setcounter{equation}{0}\setcounter{footnote}{0}

In the following, the explicit expressions for some special cases
of scalar 2-, 3- and 4-point integrals and of photon phase space integrals
will be provided, which
have been derived in course of the calculation of 
the photonic corrections usually developing IR and/or
on-shell singularities. The dimensional regularisation enables the 
extraction of the UV-divergence occurring in the scalar and vectorial
2-point integrals $B_{0,1}$ 
($\int_D\equiv\mu^{4-D} \, \int \frac{\textstyle{d^D k}}
{\textstyle{(2 \pi)^D}}$)
\begin{equation}
\frac{i}{16 \pi^2} (B_0 ; p_{\mu}\, B_1)(p^2,m_1,m_2) =
\mu^{4-D} \, \int \frac{d^D k}{(2 \pi)^D} \,
\frac{(1 ; k_{\mu})}{[k^2-m_1^2]\; [(k+p)^2-m_2^2]} \; ,
\end{equation}
so that they can be written as follows \cite{holl}:
\begin{eqnarray}
B_0(p^2,m_1,m_2) & = & \Delta -\int_0^1
dx \; \log\frac{x^2 p^2-x(p^2+m_1^2-m_2^2)+m_1^2-i\eps}{\mu^2}\\
B_1(p^2,m_1,m_2) & = & \frac{1}{2 p^2}\, [m_1^2(\Delta_{m_1}+1)
-m_2^2(\Delta_{m_2}+1)\nonumber\\
& & +(m_2^2-m_1^2-p^2)\, B_0(p^2,m_1,m_2)]\; .
\end{eqnarray}

The following results for the scalar integrals have been used
\cite{holl}, \cite{denner}:   
\begin{equation}
B_0(p^2,\lambda=0,m) = \Delta_m+2
+\left(\frac{m^2}{p^2}-1\right)
\; \log\left(1-\frac{p^2}{m^2}-i\eps\right)
\end{equation}
\begin{equation}
\left.\frac{\partial B_0(p^2,\lambda,m)}{\partial p^2}\right|_{p^2=m^2}
 =  -\frac{1}{m^2}\, \left[\lo{\lambda}{m}\right)+1\right]
\end{equation}
\begin{eqnarray}
\lefteqn{C_0(s,\mf,\mfs,\lambda)= 
\int_{D=4} \frac{1}{[k^2-\lambda^2]\, [(k+\pfs)^2-\mfs^2] \, 
[(k-\pf)^2-\mf^2]}}
\nonumber\\
& & = -\frac{1}{s}\, \left[
\lo{s}{\mf\mfs}\right)\lo{\lambda^2}{s}\right)
+\frac{1}{4}\lot{s}{\mf^2}\right)+\frac{1}{4}\lot{s}{\mfs^2}\right)
+\frac{2}{3}\pi^2-i \pi \lo{\lambda^2}{s}\right)\right]  
\nonumber \\
& & 
\end{eqnarray}
\begin{eqnarray}
\lefteqn{C_0(s,M_W,\mf,\lambda) =  
\int_{D=4} \frac{1}{[k^2-\lambda^2]\, [(k-\pf)^2-\mf^2] \, 
[(k-q)^2-M_W^2]}}
\nonumber\\
& & = \frac{1}{s} \; \left[\lo{s}{\mf^2}\right)
 \log\left(1-\frac{s}{M_W^2}-i\eps\right) -\mbox{Sp}(1-\frac{M_W^2}{s})
-\frac{1}{2}\lot{M_W^2}{s}\right)-\frac{\pi^2}{6}\right]
\nonumber\\
& &
\end{eqnarray}
\begin{equation}
C_0(s=M_W^2,M_W,\mf,\lambda) = 
\frac{1}{M_W^2} \; \left[2 \lo{M_W}{\mf}\right)
\lo{\lambda}{M_W}\right)+\lot{M_W}{\mf}\right)\right]
\end{equation}
\begin{eqnarray}
C_0(1)\equiv C_0(t,\mf,\mi,M_W) &= &\int_{D=4} \frac{1}{[(k-\pf)^2-\mf^2] \, 
[(k-\pii)^2-\mi^2]\, [(k-q)^2-M_W^2]}
\nonumber\\
& = & -\frac{1}{t}\left[\mbox{Sp}(1+\frac{t+i \eps}{M_W^2})
-\frac{\pi^2}{6}\right]
\end{eqnarray}
\begin{equation}
C_0(3;4)\equiv C_0(s,M_W,(\mf;\mi),\lambda)
\end{equation}
\begin{eqnarray}
\lefteqn{C_0(2)\equiv C_0(t,\mf,\mi,\lambda) =
\int_{D=4} \frac{1}{[k^2-\lambda^2]\, [(k-\pf)^2-\mf^2] \, 
[(k-\pii)^2-\mi^2]}}
\nonumber\\
&  & =  -\frac{1}{2t}\, \left[
\lo{t^2}{\mf^2\mi^2}\right)\lo{\lambda^2}{s}\right)
-\frac{1}{4}\lot{t^2}{s^2}\right)
+\frac{1}{2}\lot{s}{\mf^2}\right)+\frac{1}{2}\lot{s}{\mi^2}\right)
+\frac{\pi^2}{3}\right]
\nonumber\\
& & 
\end{eqnarray}
\begin{eqnarray}
\lefteqn{D_0(s,t,\mf,\mi,M_W,\lambda) = \int_{D=4}
\frac{1}{[k^2-\lambda^2]\, [(k-\pf)^2-\mf^2] \, 
[(k-\pii)^2-\mi^2]\, [(k-q)^2-M_W^2]}}
\nonumber\\
& &=  -\frac{1}{t}\, \frac{1}{s-M_W^2}\, \left[
\lo{t^2}{\mf^2\mi^2}\right)\lo{M_W \lambda}{M_W^2-s-i\eps}\right)
+\lot{\mf}{M_W}\right)+\lot{\mi}{M_W}\right)
\right.\nonumber\\
& &\left.  +\mbox{Sp}(1+\frac{M_W^2}{t+i\eps})
+\frac{\pi^2}{3}\right]
\end{eqnarray}

In addition, the following soft photon phase space integrations
have been performed
($\int_k\equiv\int \frac{\textstyle{d^3 k}}
{\textstyle{2 (2\pi)^3 \, k^0}}$ and $\dw=s-M_W^2$ is considered to be
complex):
\begin{eqnarray}
\int_k \frac{\dw \,  |2 p_i p_j|}{(\dw-2 k^0 q^0)(k p_i)(k p_j)}
& \stackrel{p_i\neq p_j}{=} &
\frac{1}{2 (2\pi)^2} \, \left\{
2 \lo{(2p_i p_j)^2}{p_i^2 p_j^2}\right) \log\left(\frac{2\Delta E}{\lambda}
\frac{\dw}{\dw-2 \sqrt{s}\Delta E}\right) - I_x \right\}
\nonumber\\
& & \mbox{}
\end{eqnarray}
with 
\begin{eqnarray}
I_x & = & \lo{(2p_i p_j)^2}{p_i^2 p_j^2}\right)
\lo{s}{|2 p_i p_j|}\right)+\frac{1}{2} \lot{p_i^2}{|2 p_i p_j|}\right)
+\frac{1}{2} \lot{p_j^2}{|2 p_i p_j|}\right)
\nonumber\\
&+& 2 \mbox{Sp}(1-\frac{s}{2 p_i p_j})+\lot{s}{|2 p_i p_j|}\right)
+\left\{\frac{2 \pi^2}{3}; \frac{\pi^2}{3} \right\} \; ,
\end{eqnarray}
where the second term in the curly bracket 
has to be used, when one of the momenta $p_i,p_j$ is equal to
the CM momentum $q$.
\begin{eqnarray}
\int_k \frac{\dw\;  p^2}{(\dw-2 k^0 q^0) (kp)^2}
& = &  \frac{1}{2 (2\pi)^2} \left\{
2 \log\left(\frac{2\Delta E}{\lambda}
\frac{\dw}{\dw-2 \sqrt{s}\Delta E}\right)-\tilde{I}_x\right\}
\end{eqnarray}
with
\begin{equation}
\tilde{I}_x = \lo{s}{p^2}\right)+\{0;2\} \; ,
\end{equation}
where again the second term in the curly bracket has to be taken, when
$p\equiv q$ holds.
\begin{eqnarray}
\int_k \frac{2 p_i p_j}{(k p_i)(k p_j)}
& \stackrel{p_i\neq p_j}{=}& \frac{1}{2 (2\pi)^2} \left\{
2 \lo{(2 p_i p_j)^2}{m_i^2 m_j^2}\right)
\lo{2\Delta E}{\lambda}\right)-I_x \right\}
\end{eqnarray}
\begin{eqnarray}
\int_k \frac{p^2}{(kp)^2}
& = &  \frac{1}{2 (2\pi)^2} \left\{
2 \lo{2\Delta E}{\lambda}\right)-\tilde{I}_x\right\} \; .
\end{eqnarray}

Finally, the functions $\alpha_i$, $v_{ij}$ and $\eta_{ij}$ used in order to 
describe the $\xi_i$-dependence of the form factors
are defined as follows \cite{sir}:
\begin{equation}
v_{ij}(q^2) \equiv  \alpha_i-2 \beta_{ij}(q^2)-q^2 \eta_{ij}(q^2)
\end{equation}
with
\begin{eqnarray}
\label{eichfkt}
\frac{i}{16\pi^2}\,\alpha_{i}& = &
\, \int_D \frac{1}{[k^2-m_i^2]\, [k^2-\xi_i m_i^2]}
\nonumber\\
\frac{i}{16\pi^2}\,\beta_{ij}(q^2)& = & t^{\mu\nu}
\, \int_D \frac{k_{\mu} k_{\nu}-g_{\mu\nu} m_j^2}{[k^2-m_i^2]
\, [k^2-\xi_i m_i^2]\, [(k+q)^2-m_j^2]}
\nonumber\\
\frac{i}{16\pi^2}\,\eta_{ij}(q^2)& = & t^{\mu\nu}
\, \int_D \frac{1}{[k^2-m_i^2]
\, [k^2-\xi_i m_i^2]\, [(k+q)^2-m_j^2]}\; 
\left[2 g_{\mu\nu}+\frac{(\xi_j-1) k_{\mu}k_{\nu}}
{[(k+q)^2-\xi_j m_j^2]}\right] \; ,
\nonumber\\
& & \mbox{}
\end{eqnarray}
where the abbreviations $t^{\mu\nu}=(g^{\mu\nu}-q^{\mu}q^{\nu}/q^2)/(D-1)$
and $\int_D\equiv\mu^{4-D} \, \int \frac{\textstyle{d^D k}}
{\textstyle{(2 \pi)^D}}$ have been used.

\end{appendix}


\begin{thebibliography}{61}

\newcommand{\artref}[4]{{\sc #1}, {\it #2} {\bf #3} #4}
\newcommand{\bookref}[2]{{\sc #1}, #2}

\bibitem{sm}
\bookref
{A.Salam}{in {\em Elementary Particle Theory}, 8th Nobelsymposium,
Wiley N.Y. 1969}  \\
\artref
{S.Weinberg}{Phys.~Rev.~Lett.}{19}{(1967), 1264}\\
\artref
{S.L.Glashow}{Nucl.~Phys.}{22}{(1961), 579}\\
\artref
{S.L.Glashow,J.Iliopoulos, L.Maiani}{Phys.~Rev.}{D2}{(1970), 1285}

\bibitem{hig}
\artref
{P.W.Higgs}{Phys.~Lett.}{12}{(1964), 131} \\
\artref
{P.W.Higgs}{Phys.~Rev.~Lett.}{13}{(1964), 508} \\
\artref
{P.W.Higgs}{Phys.~Rev.~Lett.}{145}{(1966), 1156} \\
\artref
{T.W.B.Kibble}{Phys.~Rev.}{155}{(1967), 1554} \\
\artref
{R.Brout, F.Englert}{Phys.~Rev.~Lett.}{13}{(1964), 321}

\bibitem{tgc}
\bookref
{Report of the {\em Triple gauge boson couplings} working group}
{to appear in {\em Physics at LEP2}, G.Altarelli, T.Sjostrand, F.Zwirner
(eds.), CERN Report (1996)}

\bibitem{mwav}
\bookref
{M.Demarteau, for the D0 Collaboration}{talk presented at Fermilab,
March 1996}

\bibitem{mwac}
\bookref
{Report of the {\em Determination of the Mass of the W Boson} working group}
{to appear in {\em Physics at LEP2}, G.Altarelli, T.Sjostrand, F.Zwirner
(eds.), CERN Report (1996)}

\bibitem{mwacc}
\bookref
{Report of the {\em tev2000 Study Group}}{D.Amidei, R.Brock (eds.),
FERMILAB-Pub-96/082 (1996)}

\bibitem{wwcross}
\bookref
{Report of the {\em WW Cross Sections and Distributions} working group}
{to appear in {\em Physics at LEP2}, G.Altarelli, T.Sjostrand, F.Zwirner
(eds.), CERN Report (1996)}

\bibitem{cern}
\bookref
{G.Altarelli, R.Kleiss, C.Verzegnassi (eds.)}{{\em Z Physics at LEP-I},
CERN 89-08 (1989)}   

\bibitem{topfit}
\bookref
{The LEP Collaborations ALEPH, DELPHI, L3, OPAL and the
LEP Electroweak Working Group}{CERN-PPE-95-172 (1995)} and references therein 

\bibitem{bcern}
\bookref
{F.A.Berends}{{\em Z Line Shape}, published in \cite{cern}}

\bibitem{yfs}
\artref
{D.R.Yennie, S.C.Frautschi, H.Suura}{Ann.~Phys.}{13}{(1961), 379}

\bibitem{width}
\artref
{A.Denner, T.Sack}{Z.~Phys.}{C46}{(1990), 653} and references therein

\bibitem{koba}
\artref
{Particle Data Group}{Phys.~Rev.}{D50}{(1994), no.~3} 

\bibitem{spiess}
\artref
{M.B\"ohm, W.Hollik, H.Spiessberger}{Fort.~Phys.}{34}{no.~11 (1986), 687} 

\bibitem{br}
\artref
{F.Bloch, H.Nordsiek}{Phys.~Rev.~}{52}{(1937) 54}

\bibitem{dim}
\artref
{C.Bollini, J.Giambiagi}{Nuov.~Cim.}{12B}{(1972), 20}\\
\artref
{G.'t Hooft, M.Veltman}{Nucl.~Phys.}{B44}{(1972), 189}

\bibitem{kln}
\artref
{T.Kinoshita}{J.~Math.~Phys.}{3}{(1962) 650} \\
\artref
{T.D.Lee, M.Nauenberg}{Phys.~Rev.}{133}{(1964), 1549}

\bibitem{brb}
\artref
{F.A.Berends, R.Kleiss}{Z.~Phys.}{C27}{(1985), 365}

\bibitem{been}
\bookref
{W.Beenakker, F.A.Berends, W.L.~van Neerven}{in {\em Radiative Corrections for
$e^+ e^-$ Collisions}, Proceedings of the Ringberg Workshop 1989, 
J.H.K\"uhn (ed.), 1989}

\bibitem{dilog}
\bookref
{L.Lewin}{{\em Dilogarithms and Associated Functions}, MacDonald
London 1958}

\bibitem{jeg}
\artref
{S.Eidelmann, F.Jegerlehner}{Z.~Phys.}{C67}{(1995), 585}

\bibitem{holl}
\artref
{W.Hollik}{Fort.~Phys.}{38}{no.~3 (1990), 165} 

\bibitem{lepewg}
\bookref
{The LEP Electroweak Working Group}{LEPEWWG/95-01, March 1995}

\bibitem{lepwork}
\bookref
{K.G.Chetyrkin, J.H.K\"uhn, A.Kwiatkowski}{in {\em Reports of the
Working Group on Precision Calculations for the $Z$ Resonance},
D.Bardin, W.Hollik, G.Passarino (eds.), CERN 95-03 (1995)}

\bibitem{ths}
\artref
{A.Denner, T.Sack}{Nucl.~Phys.}{B347}{(1990), 203}

\bibitem{topm}
\bookref
{G.F.Tartarelli, for The CDF Collaboration}{talk presented at the
XXXIst Rencontre de Moriond, Electroweak Interactions and Unified Theories,
Les Arcs, France, March 1996 (to appear in the proceedings),
Fermilab-Conf-96/099} \\
\bookref
{H.Schellman, for The D0 Collaboration}{talk presented 
at the {\em Pheno'96} workshop, Madison, April 1996}

\bibitem{kripf}
\artref
{H.Spiesberger}{Phys.~Rev.}{D52}{(1995), 4936} \\
\artref
{J.Kripfganz, H.Perlt}{Z.~Phys.}{C41}{(1988), 319}

\bibitem{polk}
\bookref
{R.J.Eden, P.V.Landshoff, D.I.Olive, J.C.Polkinghorne}
{{\em The Analytic S-Matrix}, Cambridge University Press 1966}

\bibitem{zres}
\artref
{R.G.Stuart}{Phys.~Lett.}{B272}{(1991), 353} and references therein \\
\artref
{H.Veltman}{Z.~Phys.}{C62}{(1994), 35}

\bibitem{dys}
\artref
{F.J.Dyson}{Phys.~Rev.}{75}{(1949), 1736}

\bibitem{veltb}
\artref
{M.Veltman}{Physica}{29}{(1963), 186}

\bibitem{wetzel}
\artref
{W.Wetzel}{Nucl.~Phys.}{B227}{(1983), 1}

\bibitem{bard}
\artref
{D.Y.Bardin, A.Leike, T.Riemann, M.Sachwitz}{Phys.~Lett.}{B206}{(1988), 539}

\bibitem{sir}
\artref
{G.Degrassi, A.Sirlin}{Nucl.~Phys.}{B383}{(1992), 73}

\bibitem{weakwork}
\artref
{E.N.Argyres, W.Beenakker, G.J.~van Oldenborgh, S.Dittmaier, \\
J.H.Hoogland, R.Kleiss, C.G.Papadopoulos, 
G.Passarino}{Phys.~Lett.}{B358}{(1995), 339}

\bibitem{baur}
\artref
{U.Baur, D.Zeppenfeld}{Phys.~Rev.~Lett.}{75}{(1995), 1002} and
references therein

\bibitem{wit}
\bookref
{B.de Wit, J.Smith}{{\em Field Theory in Particle Physics},
North Holland Physics Publishing 1986} 

\bibitem{hard}
\artref
{M.Green, M.Veltman}{Nucl.~Phys.}{B169}{(1980), 137} 

\bibitem{denner}
\artref
{W.Beenakker, A.Denner}{Nucl.~Phys.}{B338}{(1990), 349} 

\end{thebibliography}
\end{document}